\documentclass{aa}

\usepackage{graphicx}
\usepackage{txfonts}
\usepackage{amsmath}
\usepackage{booktabs}
\usepackage{xcolor}
\usepackage[
    colorlinks=true,
    linkcolor=blue,
    citecolor=blue,
    urlcolor=blue
]{hyperref}
\usepackage{placeins}

\begin{document}

\title{PHAROS: Panchromatic spectral energy distributions of solar-like stars across evolutionary stages}

\author{
V. Sumida\inst{1}\corrauth{viktor.sumida@outlook.com}
\and R. Estrela\inst{2}
\and E. Mamajek\inst{2}
\and P. Chen\inst{2}
\and A. Valio\inst{1}
}

\institute{
Center for Radio Astronomy and Astrophysics Mackenzie (CRAAM),
Mackenzie Presbyterian University, Rua da Consola\c{c}\~ao, 930,
S\~ao Paulo, S\~ao Paulo, 01302-907, Brazil
\and
Jet Propulsion Laboratory, California Institute of Technology,
4800 Oak Grove Drive, Pasadena, California 91109, USA
}

\date{}

\abstract
{Panchromatic stellar spectral energy distributions (SEDs) provide the wavelength-dependent irradiation fields required for studies of exoplanet atmospheres, photochemistry, atmospheric escape, and surface-ultraviolet radiation. For terrestrial and sub-Neptune planets, the relevant boundary condition is not the bolometric flux alone, because X-ray-to-infrared spectral structure controls photolysis, ionisation, thermal balance, and upper-atmosphere evolution.}
{We present PHAROS, the Panchromatic Habitable-world Archive of Radiation from Observed Solar-like stars, a first release of fifteen component-traceable SEDs for nearby FGK stars spanning adopted ages of 0.26--6.5\,Gyr. The catalogue is designed to provide empirical and semiempirical irradiation fields for comparative studies of exoplanet atmospheres around solar-like stars.}
{We construct stellar surface-flux SEDs over 0.5--5000\,nm from archival \textit{XMM-Newton} and \textit{Chandra} high-energy constraints, semiempirical extreme-ultraviolet (EUV) reconstructions, \textit{HST} and \textit{IUE} ultraviolet spectra, Gaia DR3 XP and ancillary optical spectrophotometry, and PHOENIX photospheric extensions. When the X-ray data support spectral modelling, target-specific coronal plasma models define the high-energy segment and anchor the EUV normalisation. For targets without a robust target-specific coronal constraint, empirical high-energy templates are normalised to literature luminosity measurements.}
{The first release contains fifteen SEDs with segment-level provenance that distinguishes directly observed intervals from reconstructed components, scaled empirical templates, bridges, and model-derived extensions. The strongest diversity across the sample occurs at X-ray, EUV, and ultraviolet wavelengths. Younger age-context groups generally show stronger high-energy output relative to their photospheric continua, although the dispersion within each group demonstrates that age alone does not uniquely determine the irradiation environment.}
{PHAROS provides traceable panchromatic irradiation fields for solar-like stars while retaining the wavelength-dependent observational and modelling provenance required to assess reconstruction-dependent uncertainties. The products are intended for studies of planetary atmospheres, photochemistry, atmospheric escape, habitability, forward modelling, and atmospheric retrievals, including atmospheric-context studies in preparation for future direct-imaging facilities such as NASA's Habitable Worlds Observatory.}


\maketitle

\section{Introduction}
\label{sec:introduction}

Characterising exoplanet atmospheres requires an accurate representation of the radiation field emitted by the host star.
For terrestrial and sub-Neptune planets, the stellar spectral energy distribution (SED), rather than the bolometric irradiation alone, provides the wavelength-dependent boundary condition that controls atmospheric heating, photolysis, ionisation, and long-term atmospheric evolution. In particular, ultraviolet photons regulate the production and destruction pathways of key molecules such as H$_2$O, CO$_2$, CH$_4$, O$_2$, and O$_3$, while X-ray and extreme-ultraviolet (EUV) photons heat and ionise the upper atmosphere and can drive atmospheric escape \citep{Watson.et.al.1981Icar...48..150W, Segura.et.al.2005AsBio...5..706S, Hu.et.al.2012ApJ...761..166H, Owen.and.Jackson.2012MNRAS.425.2931O}.

Biosignature interpretation is particularly sensitive to the wavelength dependence of the host-star radiation field. The balance between far-ultraviolet (FUV) and near-ultraviolet (NUV) irradiation regulates the production and destruction pathways of oxygen- and carbon-bearing species. High FUV fluxes can enhance the photodissociation of CO$_2$ and H$_2$O, whereas weak NUV irradiation can reduce the destruction of photochemically produced O$_3$, potentially generating abiotic O$_2$/O$_3$ signatures under some atmospheric and surface boundary conditions \citep{Segura.et.al.2005AsBio...5..706S,Hu.et.al.2012ApJ...761..166H,Domagal-Goldman.et.al.2014ApJ...792...90D,Harman.2015ApJ...812..137H,Meadows.et.al.2018AsBio..18..630M}. This stellar context will be especially important for future direct-imaging facilities such as the Habitable Worlds Observatory (HWO), which is intended to detect and spectroscopically characterise potentially habitable terrestrial planets around nearby Sun-like stars \citep{Astro2020.2021pdaa.book.....N,Stark.et.al.2019JATIS...5b4009S}. The interpretation of such planetary spectra will require the atmospheric observables to be evaluated together with the irradiation environment that drives the underlying chemistry \citep{Schwieterman.et.al.2018AsBio..18..663S,Feng.et.al.2018AJ....155..200F}.

At shorter wavelengths, X-ray and extreme-ultraviolet (EUV) radiation provides a direct link between stellar magnetic activity and atmospheric evolution. These photons heat and ionise the upper atmosphere and can contribute to hydrodynamic escape, particularly for close-in planets and young or magnetically active stars \citep{Owen.and.Jackson.2012MNRAS.425.2931O,Nishioka.et.al.2023JGRA..12831405N}. The EUV region, however, is largely inaccessible for stars other than the Sun because of strong absorption by interstellar neutral hydrogen. Stellar EUV irradiance must therefore generally be reconstructed from indirect constraints, including X-ray observations, ultraviolet emission lines, empirical scaling relations, and coronal or transition-region models \citep{Sanz-Forcada.et.al.2011A&A...532A...6S,Youngblood.et.al.2016ApJ...824..101Y,Namekata.et.al.2023ApJ...945..147N}.

Several community efforts have addressed this need for panchromatic stellar irradiance products. Among them, the MUSCLES Treasury Survey established a benchmark framework for constructing X-ray-to-infrared SEDs of nearby low-mass exoplanet host stars by combining \textit{HST} ultraviolet observations, \textit{XMM-Newton} and \textit{Chandra} X-ray data, reconstructed Ly$\alpha$ and EUV emission, and photospheric model spectra \citep{France.et.al.2016ApJ...820...89F, Youngblood.et.al.2016ApJ...824..101Y,Loyd.et.al.2016ApJ...824..102L,Youngblood.et.al.2017ApJ...843...31Y,Loyd.et.al.2018ApJ...867...71L,Wilson.et.al.2025ApJ...978...85W}.
These surveys demonstrated the value of traceable, community-ready stellar SED products for exoplanet atmospheric modelling, while also showing that ultraviolet and X-ray fluxes can vary substantially even among stars with similar fundamental parameters.

Motivated by the need for comparable stellar irradiation products for solar-like stars, we present PHAROS\footnote{\url{https://github.com/ViktorSumida/PHAROS-catalog}}, the Panchromatic Habitable-world Archive of Radiation from Observed Solar-like stars, a catalogue of hybrid panchromatic SEDs designed for exoplanet atmospheres, photochemistry, climate, and habitability studies around solar-like stars. PHAROS focuses on FGK stars spanning different activity levels and evolutionary stages, providing empirical and semiempirical stellar irradiation fields.

A defining feature of PHAROS is its wavelength-dependent source attribution. Rather than treating each final spectrum as a uniformly observed SED, we retain the origin of each segment: coronal constraints at short wavelengths, semiempirical EUV reconstructions, archival UV and optical spectra where available, and long-wavelength photospheric extensions scaled to the observed continuum. This structure allows users to identify which regions are directly observed and which rely on reconstruction, empirical templates, or stellar-atmosphere models.

For each target, the release includes stellar-surface flux spectra, component tables, diagnostic files, uncertainty estimates where available, wavelength-coverage summaries, and standardised FITS products. These files are intended for use in one-dimensional and three-dimensional atmospheric models, photochemical networks, atmospheric escape calculations, and comparative studies of stellar irradiation environments. In this way, the catalogue provides a reproducible link between observed solar-like stellar radiation fields and planetary atmospheric modelling.

The paper is organised as follows. In Sect.~\ref{sec:data}, we describe the stellar sample and the observational data used to construct the PHAROS SEDs. In Sect.~\ref{sec:sed_construction}, we present the SED construction methodology, including the X-ray modelling, EUV reconstruction, preparation of the observed UV and optical components, photospheric extension, final stitching, uncertainty treatment, and catalogue products. In Sect.~\ref{sec:results}, we present the final panchromatic SEDs, their component provenance, and the target-specific reconstruction results for the PHAROS sample. In Sect.~\ref{sec:discussion}, we discuss the physical interpretation of the catalogue and the principal observational and methodological limitations relevant to its use. Finally, Sect.~\ref{sec:conclusions} summarises the main conclusions and the broader applications of the PHAROS products.

\section{Stellar Sample and Observational Data}
\label{sec:data}

The first release contains fifteen nearby F-, G-, and K-type stars selected to sample solar-like irradiation environments across a range of ages and magnetic activity levels. The sample is not intended to be volume-complete. Instead, it is designed to provide well-documented SEDs spanning young active solar analogues, intermediate-age stars, solar-age stars, and mature systems older than the present-day Sun. 

The observational basis of the PHAROS SEDs consists of high-energy data from \textit{XMM-Newton} and \textit{Chandra}, ultraviolet spectra from \textit{HST} and \textit{IUE} when available, optical spectrophotometry from Gaia DR3 XP, and ancillary UV/optical spectral libraries such as NGSL and CFLIB. The model components employed to construct the panchromatic SEDs, including the X-ray spectral models, EUV reconstruction, PHOENIX photospheric extension, flux scaling, and final stitching procedure, are described in Sect.~\ref{sec:sed_construction}.

\subsection{Target selection}
\label{subsec:target_selection}

Targets were chosen to be broadly solar-like FGK stars with enough ancillary information to construct a traceable panchromatic SED. In practice, this required either a usable X-ray observation or a reliable literature high-energy constraint, ultraviolet coverage from \textit{HST}, \textit{IUE}, or stellar spectral libraries when available, and optical spectrophotometry capable of anchoring the photospheric continuum and the PHOENIX extension. These criteria prioritise physical usefulness for atmospheric modelling rather than statistical completeness.

The selected sample is composed of EP~Eri, $\pi^1$~UMa, $\xi$~Boo~A, $\iota$~Hor, $\kappa^1$~Cet, $\tau$~Boo~A, $\beta$~Com, 70~Oph~A, 36~Oph~B, 18~Sco, $\upsilon$~And~A, $\alpha$~Cen~A, $\lambda$~Ser, 10~Tau, and 47~UMa. Together, these stars span adopted ages from 0.26 to 6.5\,Gyr, from young magnetically active solar analogues to mature stars older than the present-day Sun. This age coverage is particularly useful for atmospheric modelling because the high-energy output of solar-like stars evolves strongly with age and magnetic activity, while rotation provides an empirical age constraint for stars within the applicable gyrochronology domain \citep{Ribas.et.al.2005ApJ...622..680R,
Bouma.et.al.2023ApJ...947L...3B}.

\begin{figure*}[t]
\centering
    \includegraphics[width=0.99\textwidth]{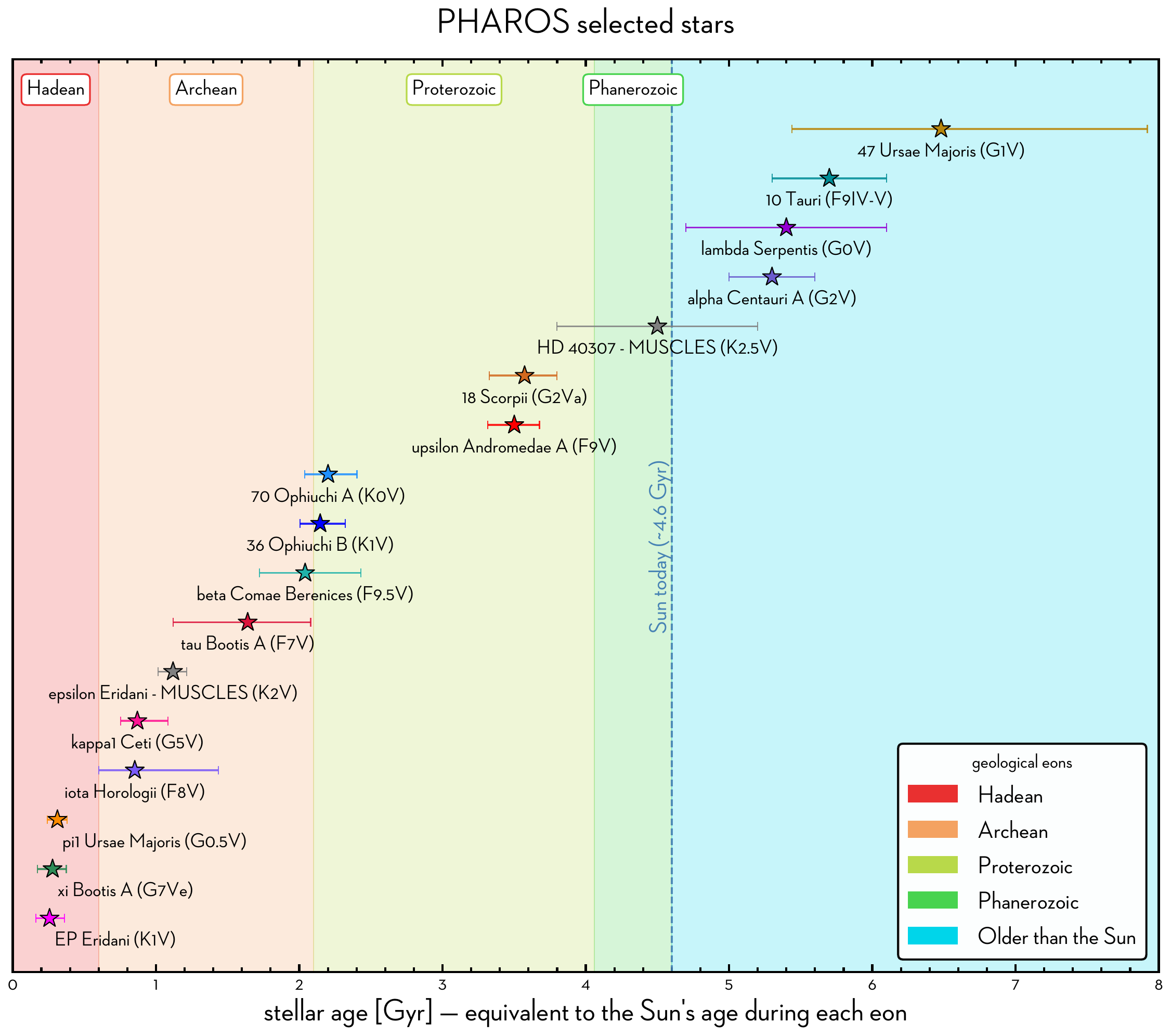}
    \caption{Age distribution of the selected PHAROS stars. The additional comparison targets, $\epsilon$~Eri and HD~40307, are FGK stars well suited to the scientific scope of PHAROS, but are excluded from the first-release sample because panchromatic SED products are already available for both stars from the MUSCLES Treasury Survey. Horizontal bars show the adopted age uncertainties for each target. The shaded regions indicate the approximate age of the Sun during the Hadean, Archean, Proterozoic, and Phanerozoic eons of Earth's history, with the dashed vertical line marking the present solar age. Geological boundaries follow the International Chronostratigraphic Chart \citep{Cohen.et.al.2013}. The adopted ages and age references for the PHAROS targets are summarised in Table~\ref{tab:stellar_parameters}. For $\epsilon$~Eri, we adopt the $1.1\pm0.1$\,Gyr age derived by \citet{Sanghi.et.al.2026AJ....171..225S}, using gyrochronology relations. For HD~40307, we adopt $4.5\pm0.7$\,Gyr from \citet{Barnes.et.al.2007ApJ...669.1167B}, where the uncertainty corresponds to the representative $\sim15\%$ gyrochronology precision quoted in that work.}
    \label{fig:pharos_age_timeline}
\end{figure*}

A comparison between the age distribution of the PHAROS targets and the age of the Sun during major geological eons on Earth is presented in Fig.~\ref{fig:pharos_age_timeline}. 
The geological-eon mapping is used only as a contextual reference for comparing stellar irradiation environments with broad epochs in Earth's history; it should not be interpreted as implying that each target is a direct evolutionary analogue of the Sun at that epoch.

\subsection{Stellar parameters}
\label{subsec:stellar_parameters}

For each PHAROS target, we compiled the stellar parameters required to convert the observed and reconstructed spectra into physically useful irradiation products. These parameters include the spectral type, effective temperature, stellar radius, stellar mass, metallicity, rotation period, age, distance, and Gaia DR3 source identifier when available. Distances are based on Gaia parallaxes whenever possible, using Gaia DR3 as the default astrometric reference \citep{Gaia.Collaboration.2023A&A...674A...1G}. Spectral types and effective temperatures were adopted from target-specific literature sources and compared, when appropriate, with empirical main-sequence dwarf calibrations \citep[e.g.,][]{Pecaut.and.Mamajek.2013ApJS..208....9P}.

The stellar radius enters directly because the catalogue reports fluxes at the stellar surface. When a directly measured radius was available, for example from interferometry, asteroseismology, or detailed stellar characterisation studies, that value was adopted. Otherwise, we used the most recent and internally consistent literature estimate available for each target. The conversion from observed flux density, $F_{\lambda,\oplus}$, to stellar-surface flux density, $F_{\lambda,\star}$, is given by
\begin{equation}
    F_{\lambda,\star} =
    F_{\lambda,\oplus}
    \left(\frac{d}{R_\star}\right)^2 ,
    \label{eq:surface_flux_scaling}
\end{equation}
where $d$ is the stellar distance, and $R_\star$ is the adopted stellar radius.

Ages were assigned using a combination of literature estimates and homogeneous rotation-based age calculations. For targets with suitable measured rotation periods and effective temperatures within the applicable domain, we calculated age posteriors using the interpolation-based gyrochronology calibration of \citet{Bouma.et.al.2023ApJ...947L...3B}. The calculations used the adopted $P_{\rm rot}$ and $T_{\rm eff}$ values together with their available measurement uncertainties.

For $\pi^1$~UMa and $\kappa^1$~Cet, the adopted rotation periods are weighted means of measurements reported by \citet{Donahue.et.al.1996ApJ...466..384D} and
\citet{Gaidos.et.al.2000AJ....120.1006G}. For $\xi$~Boo~A, $\beta$~Com, 36~Oph~B, 70~Oph~A, and 18~Sco, we adopt the rotation periods from \citet{Olspert.et.al.2018AA...619A...6O}. The latter study analysed a longer portion of the Mount Wilson Ca~{\sc\!ii} H\&K time series than the earlier period estimates of \citet{Donahue.et.al.1996ApJ...466..384D}. The two sets of periods are generally mutually consistent, but the longer temporal baseline provides improved constraints on the adopted rotation periods.
The rotation-based ages are reported as posterior medians with corresponding 16th--84th percentile intervals.

For the remaining targets, we adopted literature ages derived using methods appropriate to each system, including evolutionary, isochrone, and asteroseismic analyses. In particular, gyrochronology provides only a lower limit for $\lambda$~Ser and an approximate age for 47~UMa; we therefore retain the more informative independent literature determinations. The adopted ages, uncertainties, and methodologies are summarised in Table~\ref{tab:stellar_parameters}.

Because age determinations for field FGK stars can be method-dependent and affected by rotation-period variability, differential rotation, activity cycles, and calibration systematics, the tabulated uncertainties should be interpreted as adopted working uncertainties rather than a complete systematic error budget. This distinction is important because the high-energy emission of solar-like stars depends strongly on age, rotation, and magnetic activity \citep{Ribas.et.al.2005ApJ...622..680R,
Bouma.et.al.2023ApJ...947L...3B}.

\begin{table*}[t]
\caption{Adopted stellar parameters for the PHAROS sample.}
\label{tab:stellar_parameters}
\centering
\scriptsize
\setlength{\tabcolsep}{3.2pt}
\begin{tabular}{lllclclclc}
\hline\hline
Target & HD & SpT & $R_\star$ & Ref. & $T_{\rm eff}$ & Ref. & $P_{\rm rot}$ & Ref. & Age \\
 &  &  & ($R_\odot$) &  & (K) &  & (d) &  & (Gyr) \\
\hline
EP Eri          & HD 17925    & K1V      & $0.85^{+0.04}_{-0.05}$ & Fol18  & $5199 \pm 20$ & Sou22  & $6.76 \pm 0.11$ & Don96  & $0.26 \pm 0.10$\tablefootmark{a} \\
$\xi$ Boo A     & HD 131156 A & G7Ve     & $0.863 \pm 0.011$ & Boy12a & $5458 \pm 73$ & Ram13  & $6.000 \pm 0.036$ & Ols18 & $0.28 \pm 0.10$\tablefootmark{a} \\
$\pi^1$ UMa     & HD 72905    & G0.5V    & $0.96 \pm 0.07$ & Fol16  & $5893 \pm 14$ & Sou22  & $4.748 \pm 0.045$ & Don96,Gai00 & $0.31 \pm 0.07$\tablefootmark{a} \\
$\iota$ Hor     & HD 17051    & F8V      & $1.17 \pm 0.04$ & Fuh17  & $6000 \pm 50$ & San19  & $8.20$ & San19 & $0.85^{+0.58}_{-0.25}$\tablefootmark{a} \\
$\kappa^1$ Cet  & HD 20630    & G5V      & $0.919 \pm 0.025$ & Boy12a & $5786 \pm 87$ & Sou24  & $9.211 \pm 0.045$ & Don96,Gai00 & $0.87^{+0.21}_{-0.12}$\tablefootmark{a} \\
$\tau$ Boo A    & HD 120136   & F7V      & $1.440 \pm 0.031$ & Ros21  & $6387 \pm 44$ & Tak07  & -- & -- & $1.64^{+0.44}_{-0.52}$\tablefootmark{b} \\
$\beta$ Com     & HD 114710   & F9.5V    & $1.106 \pm 0.011$ & Boy12a & $5930 \pm 30$ & Sou24  & $11.99 \pm 0.10$ & Ols18 & $2.04^{+0.39}_{-0.32}$\tablefootmark{a} \\
36 Oph B        & HD 155885   & K1V      & $0.76 \pm 0.10$ & Jus20  & $5144 \pm 31$ & Sou22  & $18.87 \pm 0.42$ & Ols18 & $2.15^{+0.17}_{-0.14}$\tablefootmark{a} \\
70 Oph A        & HD 165341 A & K0V      & $0.8310 \pm 0.0044$ & Boy12b & $5298 \pm 32$ & Sou22 & $19.33 \pm 0.31$ & Ols18 & $2.20^{+0.20}_{-0.16}$\tablefootmark{a} \\
$\upsilon$ And A  & HD 9826     & F9V      & $1.631 \pm 0.014$ & Bai08 & $6154 \pm 10$ & Sou22 & $14.00$ & Hen00 & $3.50^{+0.18}_{-0.18}$\tablefootmark{a}\tablefootmark{c} \\
18 Sco          & HD 146233   & G2Va     & $1.010 \pm 0.009$ & Baz11 & $5824 \pm 30$ & Sou24 & $22.62 \pm 0.57$ & Ols18 & $3.57^{+0.22}_{-0.25}$\tablefootmark{a} \\
$\alpha$ Cen A  & HD 128620   & G2V      & $1.2234 \pm 0.0053$ & Ker17 & $5795 \pm 19$ & Ker17 & -- & -- & $5.3 \pm 0.3$\tablefootmark{d} \\
$\lambda$ Ser   & HD 141004   & G0V      & $1.363 \pm 0.031$ & Met23 & $5901 \pm 78$ & Met23 & -- & -- & $5.4 \pm 0.7$\tablefootmark{e} \\
10 Tau          & HD 22484    & F9IV--V  & $1.622 \pm 0.024$ & Boy12a & $5997 \pm 44$ & Boy12a & -- & -- & $5.7 \pm 0.4$\tablefootmark{f} \\
47 UMa          & HD 95128    & G1V      & $1.137 \pm 0.027$ & Tak07 & $5829 \pm 95$ & Tak07 & -- & -- & $6.48^{+1.44}_{-1.04}$\tablefootmark{g} \\
\hline
\end{tabular}
\tablefoot{
The columns labelled Ref. give the source of the immediately preceding stellar parameter. Rotation periods are tabulated for targets whose adopted ages were calculated using gyrochronology, as indicated by note~\tablefootmark{a}. For $\pi^1$~UMa and $\kappa^1$~Cet, the tabulated periods are weighted means of the values reported by \citet{Donahue.et.al.1996ApJ...466..384D} and \citet{Gaidos.et.al.2000AJ....120.1006G}. For $\xi$~Boo~A, $\beta$~Com, 36~Oph~B, 70~Oph~A, and 18~Sco, we adopt the periods from \citet{Olspert.et.al.2018AA...619A...6O}, based on the longer Mount Wilson activity time series. Entries with unavailable rotation-period information identify targets whose adopted ages were obtained independently of rotation; they do not necessarily imply that no rotation-period estimate exists in the literature.\\
\tablefoottext{a}{Rotation-based ages were calculated using the interpolation-based gyrochronology calibration of \citet{Bouma.et.al.2023ApJ...947L...3B}, based on the adopted $P_{\rm rot}$ and $T_{\rm eff}$ values and their available uncertainties, as described in Sect.~\ref{subsec:stellar_parameters}. The reported values are posterior medians with corresponding 16th--84th percentile intervals.}
\tablefoottext{b}{For $\tau$ Boo A, we adopt the evolutionary-model age from \citet{Takeda.et.al.2007ApJS..168..297T} rather than deriving a rotation-based age for this target.}
\tablefoottext{c}{For $\upsilon$ And A, the rotation-based calculation uses the 14-day rotation period from \citet{Henry.et.al.2000ApJ...531..415H}, which was reported as a low-confidence rotation determination. The star also lies close to the upper effective-temperature boundary of the gyrochronology calibration, and its age should therefore be interpreted cautiously.}
\tablefoottext{d}{For $\alpha$~Cen~A, we adopt the common system age obtained from the joint classical and asteroseismic modelling of $\alpha$~Cen~A and B by \citet{Joyce.and.Chaboyer.2018ApJ...864...99J}.}
\tablefoottext{e}{The age, effective temperature, and radius of $\lambda$ Ser are taken from \citet{Metcalfe.et.al.2023AJ....166..167M}. The available gyrochronology constraint provides only a lower age limit of approximately 4~Gyr and is therefore not adopted in place of the literature age.}
\tablefoottext{f}{For 10~Tau, we adopt the isochrone age $5.7 \pm 0.4$~Gyr from \citet{Boyajian.et.al.2012ApJ...746..101B}. The same work provides the adopted interferometric radius and effective temperature.}
\tablefoottext{g}{For 47 UMa, we adopt the evolutionary-model age $6.48^{+1.44}_{-1.04}$\,Gyr from \citet{Takeda.et.al.2007ApJS..168..297T}. A gyrochronology check based on the $P_{\rm rot}=24.71\pm0.33$~d estimate from \citet{Schmitt.et.al.2020AN....341..497S} gives only a rough age of approximately 4\,Gyr. Because this rotation-based estimate is not sufficiently robust for a precise age determination, it is not adopted in place of the evolutionary-model age.}\\
References: Bai08 = \citet{Baines.et.al.2008ApJ...680..728B}; Baz11 = \citet{Bazot.et.al.2011AA...526L...4B}; Boy12a = \citet{Boyajian.et.al.2012ApJ...746..101B}; Boy12b = \citet{Boyajian.et.al.2012ApJ...757..112B}; Don96 = \citet{Donahue.et.al.1996ApJ...466..384D}; Fol16 = \citet{Folsom.et.al.2016MNRAS.457..580F}; Fol18 = \citet{Folsom.et.al.2018MNRAS.474.4956F}; Fuh17 = \citet{Fuhrmann.et.al.2017ApJ...836..139F}; Gai00 = \citet{Gaidos.et.al.2000AJ....120.1006G}; Hen00 = \citet{Henry.et.al.2000ApJ...531..415H}; Jus20 = \citet{Justesen.and.Albrecht.2020AA...642A.212J}; Ker17 = \citet{Kervella.et.al.2017AA...597A.137K}; Met23 = \citet{Metcalfe.et.al.2023AJ....166..167M}; Ols18 = \citet{Olspert.et.al.2018AA...619A...6O}; Ram13 = \citet{Ramirez.et.al.2013ApJ...764...78R}; Ros21 = \citet{Rosenthal.et.al.2021ApJS..255....8R}; San19 = \citet{Sanz-Forcada.et.al.2019AA...631A..45S}; Sou22 = \citet{Soubiran.et.al.2022yCat..36630004S}; Sou24 = \citet{Soubiran.et.al.2024AA...682A.145S}; Tak07 = \citet{Takeda.et.al.2007ApJS..168..297T}.
}
\end{table*}

\subsection{\textit{XMM-Newton} and \textit{Chandra} observations}
\label{subsec:xray_observations}

The shortest wavelengths are anchored, where possible, by archival \textit{XMM-Newton} or \textit{Chandra} observations.
\textit{XMM-Newton} provides high-throughput X-ray imaging and spectroscopy through the European Photon Imaging Camera (EPIC), including the EPIC-pn and EPIC-MOS detectors \citep{Jansen.et.al.2001A&A...365L...1J, Struder.et.al.2001A&A...365L..18S, Turner.et.al.2001A&A...365L..27T}. \textit{Chandra} provides complementary high-angular-resolution X-ray observations, which are particularly valuable for resolving close visual systems or reducing contamination from nearby sources \citep{Weisskopf.et.al.2002PASP..114....1W}.

For each target, the available high-energy observations were inspected to determine whether the stellar X-ray emission could be extracted as a clean point source. When multiple observations were available, the preferred data set was selected based on source detection significance, exposure time, instrumental configuration, background level, and contamination risk. In binary or multiple systems, \textit{Chandra} observations were preferred whenever angular resolution was required to isolate the desired stellar component.

The spectral modelling of these observations and their conversion into the PHAROS X-ray component are described in Sect.~\ref{subsec:xray_modeling}.

\subsection{\textit{HST}, \textit{IUE}, Gaia, and ancillary UV/optical spectral libraries}
\label{subsec:uv_optical_data}

The ultraviolet and optical portions of the PHAROS SEDs are assembled from archival spectroscopic and spectrophotometric data. In the ultraviolet, the preferred source is direct \textit{HST} spectroscopy, particularly observations obtained with the Cosmic Origins Spectrograph (COS) and the Space Telescope Imaging Spectrograph (STIS). COS provides high-sensitivity ultraviolet spectroscopy, while STIS provides complementary ultraviolet and optical modes with stable flux calibration and broad wavelength coverage \citep{Green.et.al.2012ApJ...744...60G, Woodgate.et.al.1998PASP..110.1183W}. When available, these data are used to constrain chromospheric and transition-region features such as H\,{\sc i} Ly$\alpha$, C\,{\sc ii}, Si\,{\sc iii}, Si\,{\sc iv}, C\,{\sc iv}, He\,{\sc ii}, and Mg\,{\sc ii}.

In addition to target-specific \textit{HST} observations, PHAROS uses the \textit{HST}/STIS Next Generation Spectral Library (NGSL) when it provides useful wavelength coverage for a given target. NGSL is a low-resolution UV--optical stellar spectral library obtained with STIS modes such as G230LB, G430L, and G750L, covering approximately 1670--10250\,\AA{} \citep{Heap.and.Lindler.2007IAUS..241...95H}. When an NGSL spectrum is available for a PHAROS target, it is treated as an observed \textit{HST}/STIS spectral component and is labelled separately from target-specific COS/STIS observations in the component tables.

When \textit{HST} spectra are unavailable or do not cover a required wavelength interval, \textit{IUE} spectra are used as secondary ultraviolet constraints. The \textit{IUE} archive provides ultraviolet spectra over approximately 1150--3200\,\AA{} \citep{Boggess.et.al.1978Natur.275..372B}. However, because \textit{IUE} data can have lower sensitivity and lower spectral resolution than modern \textit{HST} observations, and because some spectral regions may be affected by background, geocoronal contamination, or low signal-to-noise ratio, \textit{IUE} segments are treated conservatively. In the PHAROS products, \textit{IUE} data are primarily used to bridge gaps, constrain broad ultraviolet continuum levels, or provide wavelength coverage when no higher-quality data exist. These regions are explicitly identified through source labels.

The optical portion of the SED is anchored whenever possible by Gaia DR3 XP spectrophotometry. Gaia DR3 provides low-resolution BP/RP spectra for a large number of sources, with processing and external calibration described by \citet{De.Angeli.et.al.2023A&A...674A...2D} and \citet{Montegriffo.et.al.2023A&A...674A...3M}. In PHAROS, Gaia XP spectra are used to constrain the observed photospheric continuum and to set the normalisation level for the transition to the long-wavelength photospheric extension. When Gaia XP spectra are unavailable, unreliable, or potentially affected by source confusion, ancillary optical libraries or broadband photometry are used as secondary constraints.

One such ancillary optical library is the Indo-US Library of Coud\'e Feed Stellar Spectra, commonly referred to as CFLIB \citep{Valdes.et.al.2004ApJS..152..251V}. CFLIB provides optical spectra for a large sample of stars over approximately 3460--9464\,\AA{} at a spectral resolution of order 1\,\AA. In PHAROS, CFLIB spectra are used only when they improve the optical coverage of a specific target or provide a useful consistency check against Gaia XP or other optical data. Because CFLIB is a ground-based optical library rather than a space-based ultraviolet data set, CFLIB segments are labelled separately from \textit{HST}, \textit{IUE}, Gaia, and NGSL components.

For every target, the final PHAROS component tables preserve the source attribution of each wavelength interval. This distinction is essential because the UV/optical region may contain a mixture of direct \textit{HST} spectroscopy, NGSL/STIS library spectra, \textit{IUE} spectra, Gaia XP spectrophotometry, CFLIB optical spectra, and model-based extensions introduced later in the SED construction procedure.

\section{Construction of the PHAROS SEDs}
\label{sec:sed_construction}

Each SED is assembled from observational, reconstructed, empirical-template, and model-atmosphere components. The procedure is designed to retain the physical origin of each segment rather than forcing the final product to appear as a uniformly observed spectrum. Source labels and diagnostic metadata are therefore carried through the high-energy, ultraviolet, optical, and infrared regions.

For targets with usable X-ray spectra, the coronal model provides both the short-wavelength spectral shape and the luminosity anchor for the EUV reconstruction. The ultraviolet and optical observations are then inspected, placed on a common wavelength and flux-density convention, and stitched to the reconstructed region. At longer wavelengths, a PHOENIX atmosphere model scaled to the observed optical continuum supplies the photospheric extension. The assembled spectrum is finally converted into the standard observed-flux and stellar-surface-flux products.

Throughout this process, we avoid applying arbitrary smoothing across physically distinct wavelength regions. Instead, discontinuities, gaps, low-signal segments, and model transitions are documented through the component tables and diagnostic files. This approach allows users to identify which parts of the SED are directly observed, reconstructed, model-dependent, or affected by source-specific limitations.

\subsection{X-ray spectral modelling}
\label{subsec:xray_modeling}

The X-ray component sets the short-wavelength luminosity scale used in the SED construction.
For targets with sufficient X-ray counts, the extracted \textit{XMM-Newton} or \textit{Chandra} spectra are modelled using optically thin thermal plasma emission models appropriate for stellar coronae. Spectral fitting is performed with XSPEC \citep{Arnaud.1996ASPC..101...17A}, using APEC plasma models to represent the coronal emission \citep{Smith.et.al.2001ApJ...556L..91S}. When an interstellar absorption component is included, we use \texttt{tbabs} \citep{Wilms.et.al.2000ApJ...542..914W}; for the nearby PHAROS targets, this column is either negligible or fixed to a very small value, so the fitted coronal temperatures and integrated unabsorbed luminosities remain the relevant quantities for the SED construction. Depending on the quality of the available data, one-, two-, or three-temperature APEC models are adopted. The final choice is based on the statistical quality of the fit, the physical plausibility of the fitted temperatures and normalisations, and the stability of the integrated X-ray luminosity. In all APEC-based cases, the individual temperature components should be interpreted phenomenologically as compact descriptions of multi-thermal coronal emission; the quantity propagated into the EUV reconstruction is the integrated unabsorbed $L_X$ over the adopted X-ray bandpass.

For targets with usable X-ray spectra, the best-fit coronal model is exported as a wavelength-dependent flux spectrum and integrated over the adopted X-ray bandpass. This model spectrum, rather than the detector-space count spectrum, is used as the X-ray segment of the PHAROS SED. In these cases, the resulting X-ray luminosity, $L_X$, provides both the normalisation of the high-energy spectral segment and the luminosity anchor used for the EUV reconstruction described in Sect.~\ref{subsec:euv_reconstruction}.

For $\lambda$\,Ser and 47\,UMa, the high-energy treatment differs from the target-specific APEC-based procedure. In both cases, the available X-ray information is too weak to define a robust target-specific coronal spectral shape for the adopted high-energy SED. Therefore, the soft-X-ray segment from 0.5 to 12.4\,nm is represented by a FISM2-based spectral template scaled to an adopted literature X-ray luminosity. Thus, for these two targets, $L_X$ still acts as the high-energy luminosity anchor, but the wavelength-dependent soft-X-ray shape is supplied by the scaled FISM2 template rather than by a target-specific APEC model.

\subsection{EUV reconstruction}
\label{subsec:euv_reconstruction}

The EUV region controls upper-atmosphere heating, ionisation, and several photochemical pathways, but it is generally inaccessible for nearby stars because of interstellar H\,{\sc i} absorption. We therefore reconstruct this interval semiempirically, using the X-ray luminosity defined in Sect.~\ref{subsec:xray_modeling} as the normalisation anchor.

For most PHAROS targets, the EUV spectral shape is based on the semiempirical XUV/FUV reconstruction framework of \citet{Namekata.et.al.2023ApJ...945..147N}. \citet{Namekata.et.al.2023ApJ...945..147N} derived wavelength-dependent solar scaling relations between the disc-integrated XUV/FUV spectrum and the total unsigned magnetic flux using Sun-as-a-star observations over 0.1--180~nm. Their reconstruction was tested against active young solar-like G dwarfs, including EK\,Dra, $\pi^1$\,UMa, and $\kappa^1$\,Cet, which makes the method particularly relevant for the young and active solar analogues included in PHAROS. In the original formulation, the stellar XUV/FUV spectrum is estimated from magnetic flux; however, homogeneous surface magnetic flux measurements are not available for the full PHAROS sample. We therefore use the Namekata-based spectrum primarily to provide the wavelength-dependent shape of the unobserved XUV/EUV region.

The absolute EUV normalisation is tied to the X-ray luminosity through the empirical relation of \citet{Sanz-Forcada.et.al.2011A&A...532A...6S}. The authors modelled the coronal emission of late-type planet-host stars observed in X-rays and calibrated a relation between the X-ray luminosity, measured over 5--100\,\AA{}, and the EUV luminosity, estimated over 100--920\,\AA{}:
\begin{equation}
    \log L_{\rm EUV} = (4.80 \pm 1.99) + (0.860 \pm 0.073)\log L_X \, ,
    \label{eq:sanz_lx_leuv}
\end{equation}
where both luminosities are given in erg~s$^{-1}$. In PHAROS, this relation is used as an X-ray-anchored EUV normalisation rather than as an age-based evolutionary prescription. The X-ray luminosity is used to estimate $L_{\rm EUV}$, and the reconstructed EUV segment is scaled so that its integrated luminosity matches the Sanz-Forcada prediction.

For the targets using the Namekata/Sanz reconstruction, the Namekata-based segment provides the EUV spectral shape and is normalised to the Sanz-Forcada $L_{\rm EUV}$.
The final scaled reconstruction fills the interval between the X-ray model and the beginning of the observed ultraviolet spectrum, typically from 12.4 to 115.0\,nm. This procedure should be interpreted as an X-ray-anchored EUV reconstruction rather than a direct EUV observation. Although the original Namekata framework relates the solar XUV/FUV spectrum to magnetic flux, PHAROS does not adopt target-specific surface magnetic flux measurements as the primary normalisation parameter for the full sample. Instead, the absolute scale of the reconstructed EUV segment is set by the available X-ray constraint.

For four PHAROS targets, $\alpha$~Cen~A, $\lambda$~Ser, 10~Tau, and 47~UMa, the EUV spectral shape is instead supplied by FISM2. For $\alpha$~Cen~A and 10~Tau, the X-ray region remains anchored by a target-specific APEC model, while FISM2 provides the spectral shape of the reconstructed EUV region. For $\lambda$~Ser and 47~UMa, the available X-ray information does not support a robust target-specific coronal spectral shape, and FISM2 is therefore also used to represent the soft-X-ray region, normalised to the adopted X-ray luminosity.

In the FISM2/Sanz implementation, the 12.4--91.2\,nm interval is scaled so that its integrated luminosity matches the EUV luminosity inferred from the Sanz-Forcada relation or adopted from the Sanz-Forcada tabulation. The region longward of 91.2\,nm is treated as a transition between the reconstructed high-energy segment and the observed ultraviolet data. In this context, FISM2 is not used as an unscaled solar analogue; it supplies the wavelength-dependent shape of the weakly constrained high-energy spectrum, while the absolute normalisation is set by the adopted stellar luminosity constraints.

In the EUV regions reconstructed using Namekata/Sanz and FISM2/Sanz, the uncertainty is treated as an analytic multiplicative envelope. The central spectral shape is kept fixed, and the uncertainty is applied to the EUV normalisation as a log-normal prediction scatter:
\begin{equation}
    F_{\lambda,p}(\lambda) =
    F_{\lambda,0}(\lambda)
    10^{z_p \sigma_{\log L_{\rm EUV}} w(\lambda)} ,
    \label{eq:euv_uncertainty}
\end{equation}
where $F_{\lambda,0}$ is the central reconstructed spectrum, $z_p$ is the standard-normal quantile associated with percentile $p$, $\sigma_{\log L_{\rm EUV}}$ is the adopted scatter in $\log_{10} L_{\rm EUV}$, and $w(\lambda)$ is a wavelength-dependent weight. We adopt $\sigma_{\log L_{\rm EUV}}=0.50$~dex as the default prediction scatter for the reconstructed EUV luminosity. The tabulated percentiles are 2.5, 16, 50, 84, and 97.5, corresponding to the 68\% and 95\% analytic uncertainty envelopes.

The wavelength-dependent weight applies the full EUV normalisation uncertainty over 12.4--91.2\,nm. Between 91.2 and 115.0\,nm, the uncertainty is smoothly tapered to zero to avoid an artificial discontinuity at the transition to the observed ultraviolet data. Redward of 115.0\,nm, no Namekata/Sanz or FISM2/Sanz uncertainty is applied, because those regions are constrained by observed or separately modelled UV/optical/infrared components. By default, the analytic EUV envelope does not modify the X-ray region below 12.4\,nm; uncertainties associated with the X-ray spectral modelling or adopted literature $L_X$ values are treated separately.

This uncertainty treatment is not a Monte Carlo propagation of all observational and modelling errors. Instead, it is a deterministic representation of the dominant systematic uncertainty associated with the reconstructed EUV normalisation. The purpose is to provide a practical range of plausible EUV levels while preserving the central SED used in the PHAROS component tables. The resulting uncertainty tables can be used to test the sensitivity of photochemical and atmospheric escape calculations to the unobserved EUV flux.

\subsection{Preparation of observed UV and optical components}
\label{subsec:uv_optical_preparation}

After the short-wavelength reconstruction is defined, the available ultraviolet and optical observations described in Sect.~\ref{subsec:uv_optical_data} are placed on a common wavelength and flux-density convention.
This step is performed on a target-by-target basis because the available archival coverage, overlap regions, and data quality vary across the sample.

All input spectra are converted to flux density per unit wavelength in consistent units and are inspected over the wavelength intervals used in the final product. Segments affected by detector-edge artefacts, very low signal, obvious discontinuities, or unreliable overlap behaviour are excluded from the stitched SED or restricted to the wavelength range where they provide useful information. When multiple spectra cover the same wavelength region, priority is given to target-specific space-based observations, followed by archival spectra or spectral-library data that provide stable continuum constraints.

For ultraviolet spectra, the preparation focuses on preserving reliable chromospheric and transition-region emission while avoiding artificial jumps between reconstructed and observed components. \textit{IUE} spectra are used primarily when they fill gaps between the reconstructed high-energy segment and the higher-quality ultraviolet or optical data. In overlap regions, multiplicative scaling factors are estimated from selected wavelength windows that avoid strong emission lines and low-sensitivity edges whenever possible.

For the optical region, the observed spectrophotometric components are used to anchor the photospheric continuum and to set the scale for the long-wavelength model extension described in Sect.~\ref{subsec:photospheric_stitching}. When NGSL, Gaia XP, or CFLIB spectra overlap, their relative consistency is evaluated before selecting the segment used in the final stitch. The goal of this step is not to force all archival spectra into a single smoothed curve, but to define a physically and observationally traceable sequence of components for each target.

\subsection{Photospheric extension and final stitching}
\label{subsec:photospheric_stitching}

The observed UV and optical data do not generally extend continuously to the 5~$\mu$m limit adopted for the catalogue. The red-optical and infrared continuum is therefore completed with PHOENIX model atmospheres \citep{Husser.et.al.2013}, selected based on the adopted effective temperature, surface gravity, and metallicity of each target when available.

The PHOENIX spectra are used only as long-wavelength photospheric extensions and are not intended to replace observed ultraviolet, chromospheric, or transition-region emissions. Before inclusion in the final SED, each PHOENIX model is converted to the same wavelength and flux-density convention as the observed components and scaled multiplicatively to match the observed optical continuum. The scaling is derived from selected overlap windows in the red optical or near-infrared where the observed spectrum is reliable and where strong spectral features or detector-edge artefacts do not dominate the comparison.

The observed component used to anchor the PHOENIX scaling depends on the available data for each target. When Gaia DR3 XP or NGSL coverage extends sufficiently far into the red optical, these spectra generally provide the primary continuum constraint. When needed, CFLIB or other optical library spectra are used as secondary constraints or consistency checks. The resulting scale factor is then applied to the PHOENIX spectrum, and only the wavelength interval beyond the adopted observed component is retained in the final stitched SED.

The final SED is assembled by applying a target-specific sequence of wavelength intervals. The high-energy region is represented by either a target-specific APEC model or a scaled FISM2-based template, depending on the available X-ray constraint. The unobserved EUV region is reconstructed with either the Namekata/Sanz or FISM2/Sanz procedure described in Sect.~\ref{subsec:euv_reconstruction}. The ultraviolet and optical regions are then filled with the prepared observed components described in Sect.~\ref{subsec:uv_optical_preparation}, followed by the scaled PHOENIX photospheric extension at longer wavelengths.

When adjacent components overlap, the final wavelength boundaries are chosen to preserve the most reliable segment in each interval. Preference is given to target-specific observations over lower-priority archival or model components, provided that the data quality and overlap behaviour are acceptable. The final stitched SED therefore represents a continuous, component-resolved spectrum in which each wavelength interval retains its original observational, reconstructed, or modelled origin.

\subsection{Catalogue products and FITS format}
\label{subsec:catalog_products}

The catalogue is distributed as FITS files containing both the stitched science-ready spectrum and the metadata needed to interpret its construction. Each file stores the wavelength-dependent stellar surface flux density, adopted stellar parameters, wavelength coverage, component structure, and reconstruction method.

The main SED extension contains the stitched spectrum in a uniform wavelength and flux-density convention. For each wavelength point, the table stores the wavelength in nm and \AA{}, the stellar surface flux density per unit wavelength, and component identifiers that distinguish observed, reconstructed, bridge, empirical-template, photospheric, and hybrid intervals.

A separate segment table summarises the wavelength limits and source class of each major component used in the final stitch. This table records, for example, whether a given interval is derived from a target-specific X-ray plasma model, a reconstructed EUV segment, an observed ultraviolet or optical spectrum, a scaled empirical high-energy template, or a photospheric model extension. The segment information is intended to provide a compact overview of the final component sequence without requiring the user to reconstruct it from the full wavelength table.

The FITS products also include provenance and diagnostic extensions. The provenance information records the major input files, models, and literature constraints used in constructing each SED. The diagnostic information stores the key checks performed during the assembly process, including wavelength coverage, component boundaries, scaling choices, and luminosity-normalisation checks for the reconstructed high-energy regions. When an analytic EUV uncertainty envelope is available, it is stored in a method-specific uncertainty extension corresponding to the reconstruction used for that target.

\section{Results}
\label{sec:results}

The first PHAROS release comprises fifteen panchromatic stellar SEDs spanning a common wavelength range of 0.5--5000\,nm but constructed from target-dependent combinations of observational, reconstructed, empirical-template, and model components. In this section, we first examine the ensemble properties of the final spectra and their wavelength-dependent provenance, with particular attention to the variation in high-energy irradiation across the adopted age-context groups. We then present the target-specific reconstruction results, documenting the observational anchors, modelling choices, stitching procedures, and principal caveats that define each individual SED.

\subsection{Panchromatic SEDs and component provenance}
\label{subsec:pharos_seds}

\begin{figure*}
\centering
    \includegraphics[width=0.87\textwidth]{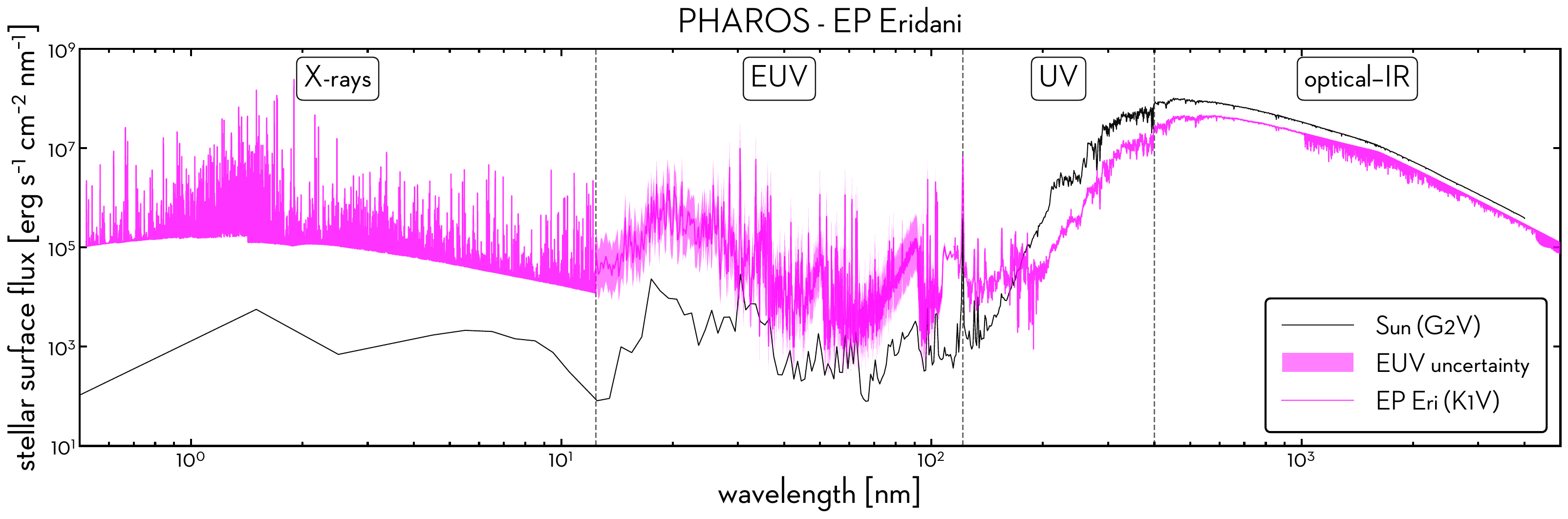}
    \includegraphics[width=0.87\textwidth]{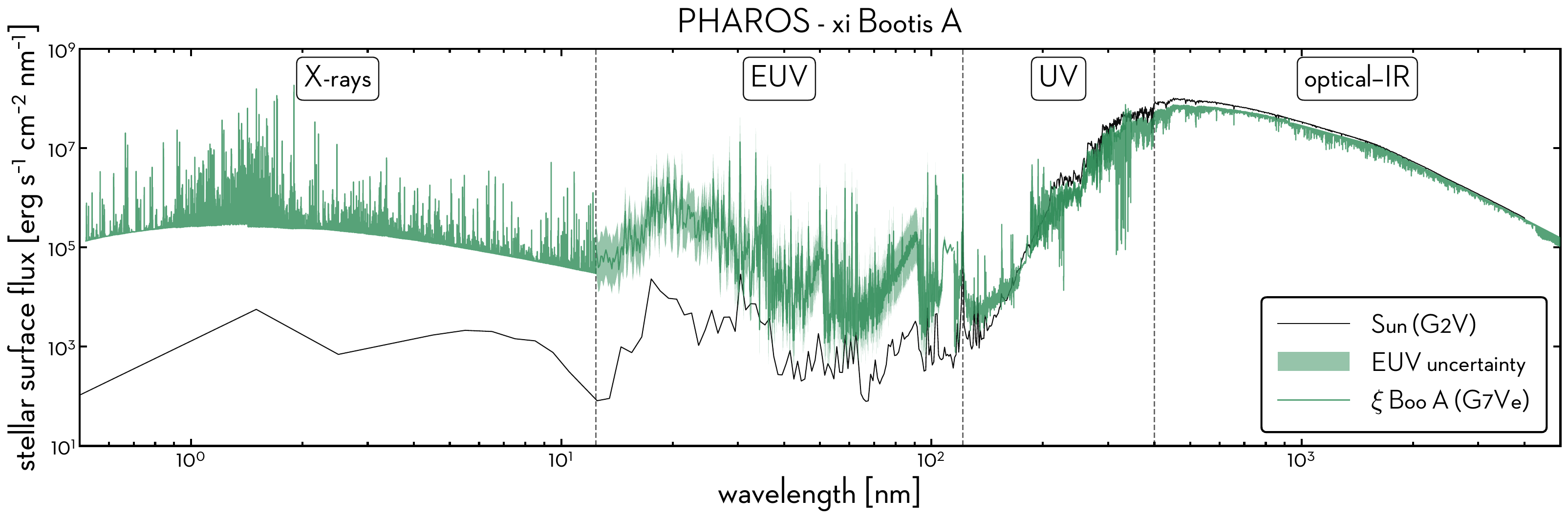}
    \includegraphics[width=0.87\textwidth]{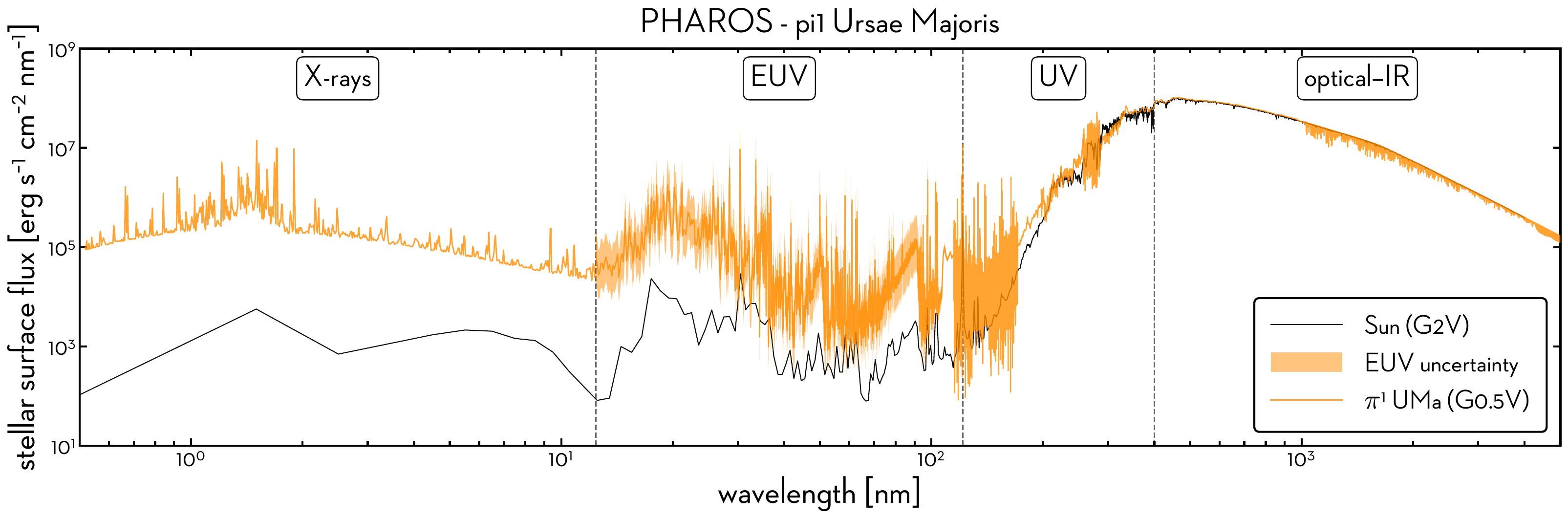}
    \caption{Final PHAROS stellar surface-flux SEDs for the Hadean-context group: EP~Eri, $\xi$~Boo~A, and $\pi^1$~UMa. Each panel shows the PHAROS SED over 0.5--5000\,nm together with the reference solar spectrum \citep{Gueymard.2004SoEn...76..423G}. The geological designation provides an age-context comparison with the early Sun and does not imply that the individual targets are exact solar evolutionary analogues.}
\label{fig:pharos_seds_hadean}
\end{figure*}

\begin{figure*}
\centering
    \includegraphics[width=0.87\textwidth]{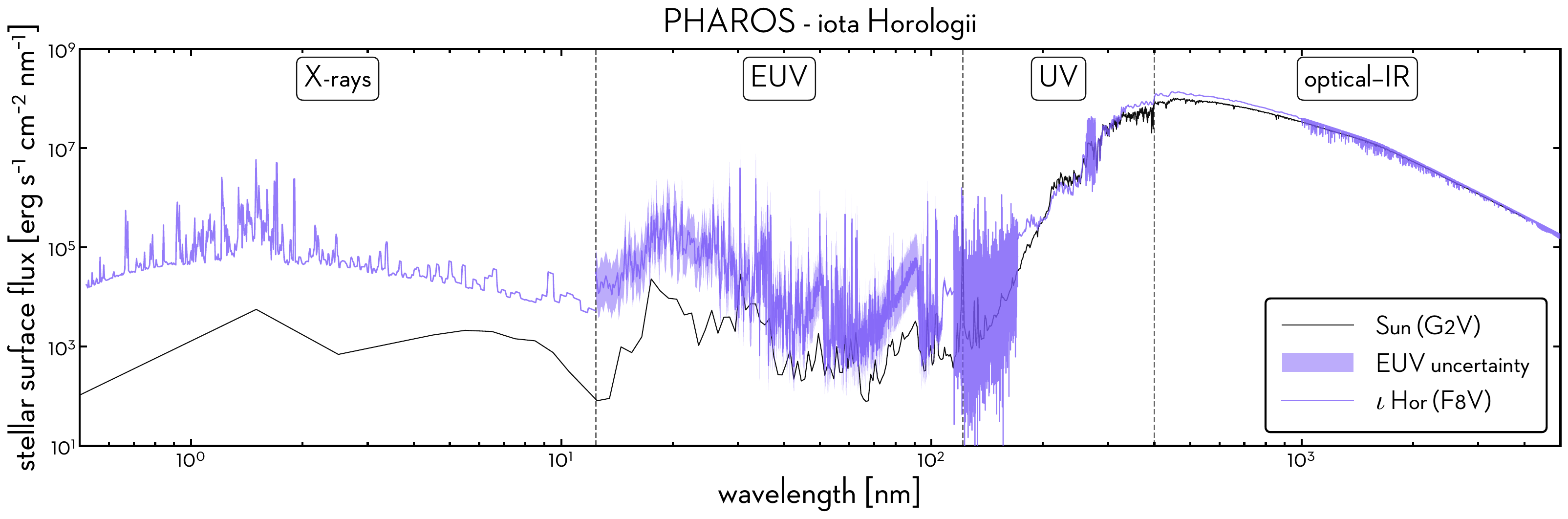}
    \includegraphics[width=0.87\textwidth]{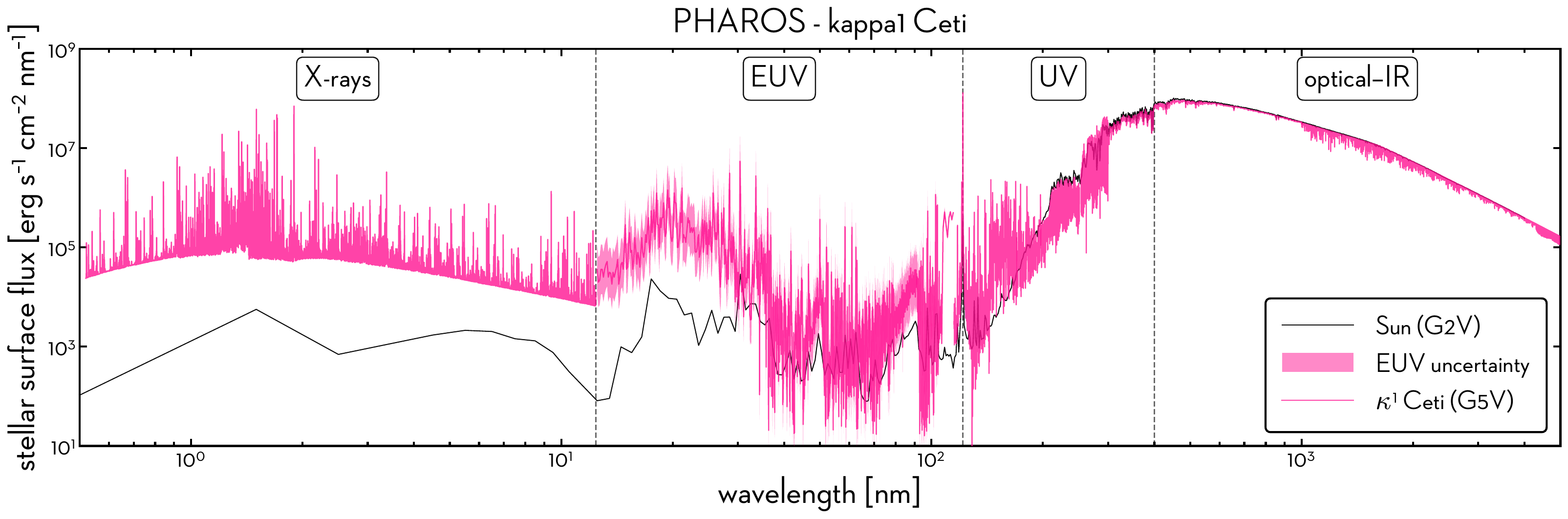}
    \includegraphics[width=0.87\textwidth]{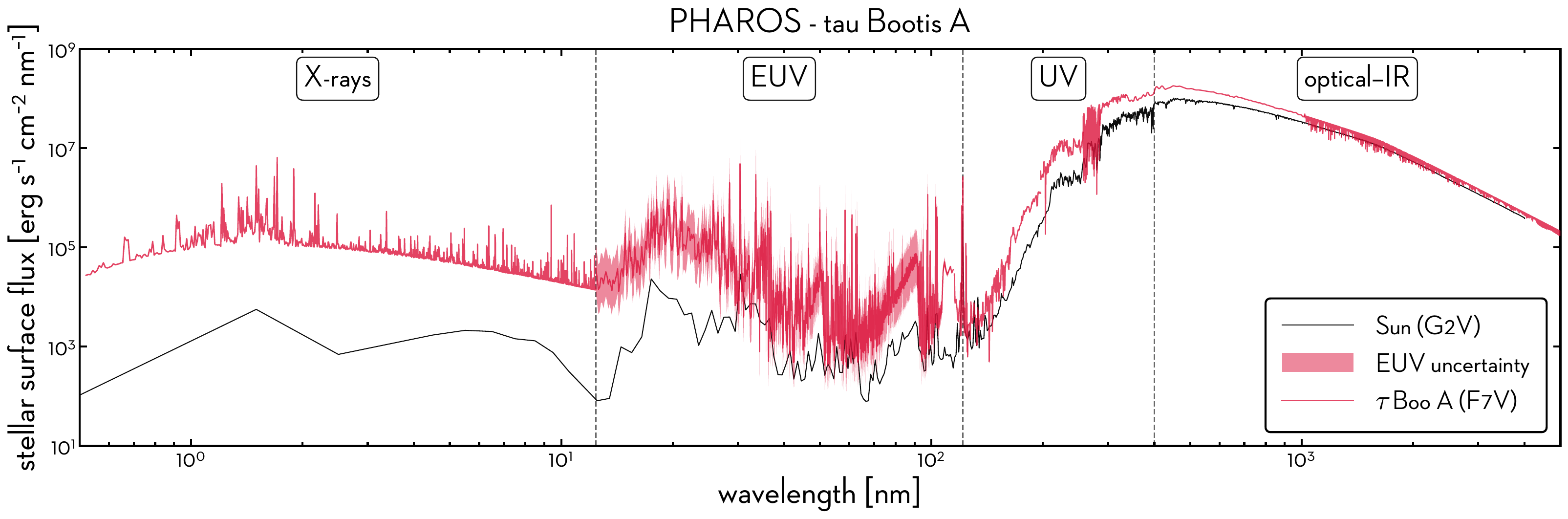}
    \includegraphics[width=0.87\textwidth]{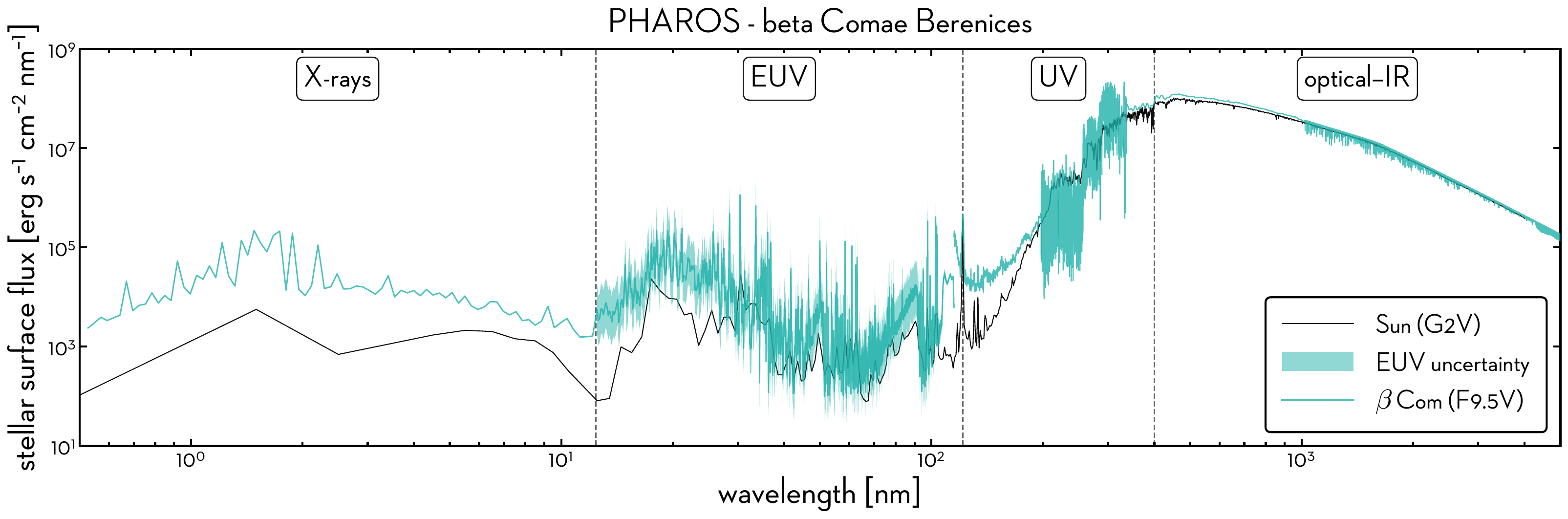}
    \caption{Final PHAROS stellar surface-flux SEDs for the Archean-context: $\iota$~Hor, $\kappa^1$~Cet, $\tau$~Boo~A, and $\beta$~Com. The spectra are shown over 0.5--5000\,nm and compared with the reference solar spectrum \citep{Gueymard.2004SoEn...76..423G} using the same plotting convention as in Fig.~\ref{fig:pharos_seds_hadean}.}
    \label{fig:pharos_seds_archean}
\end{figure*}

\begin{figure*}
\centering
    \includegraphics[width=0.87\textwidth]{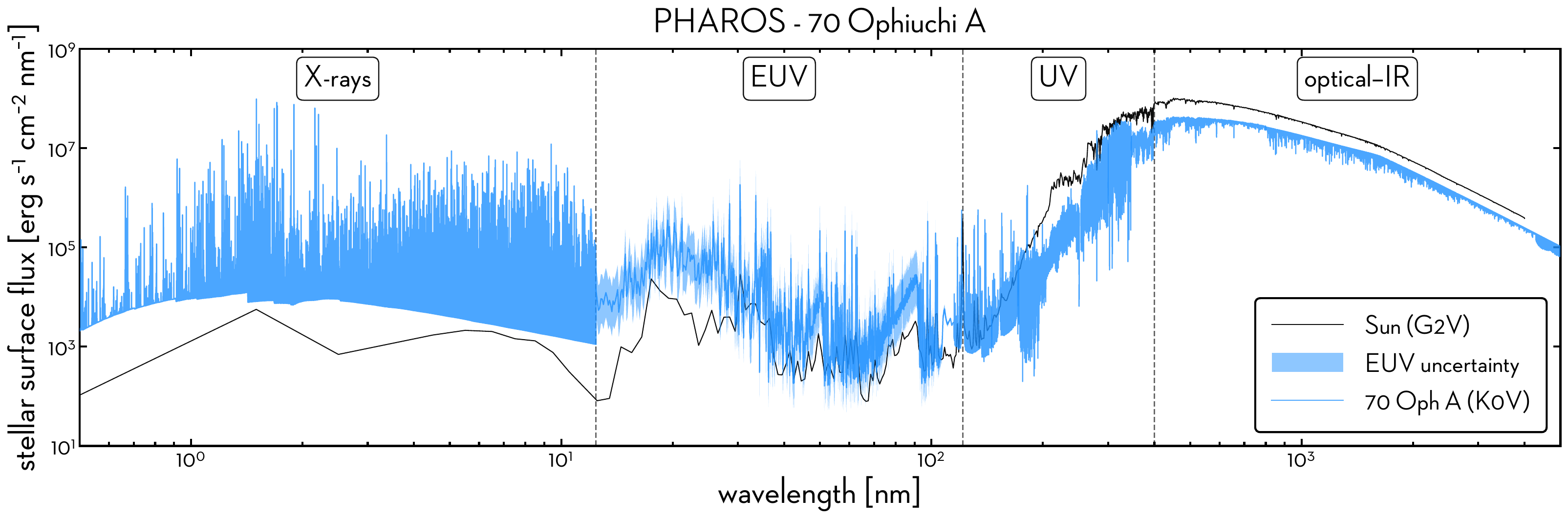}
    \includegraphics[width=0.87\textwidth]{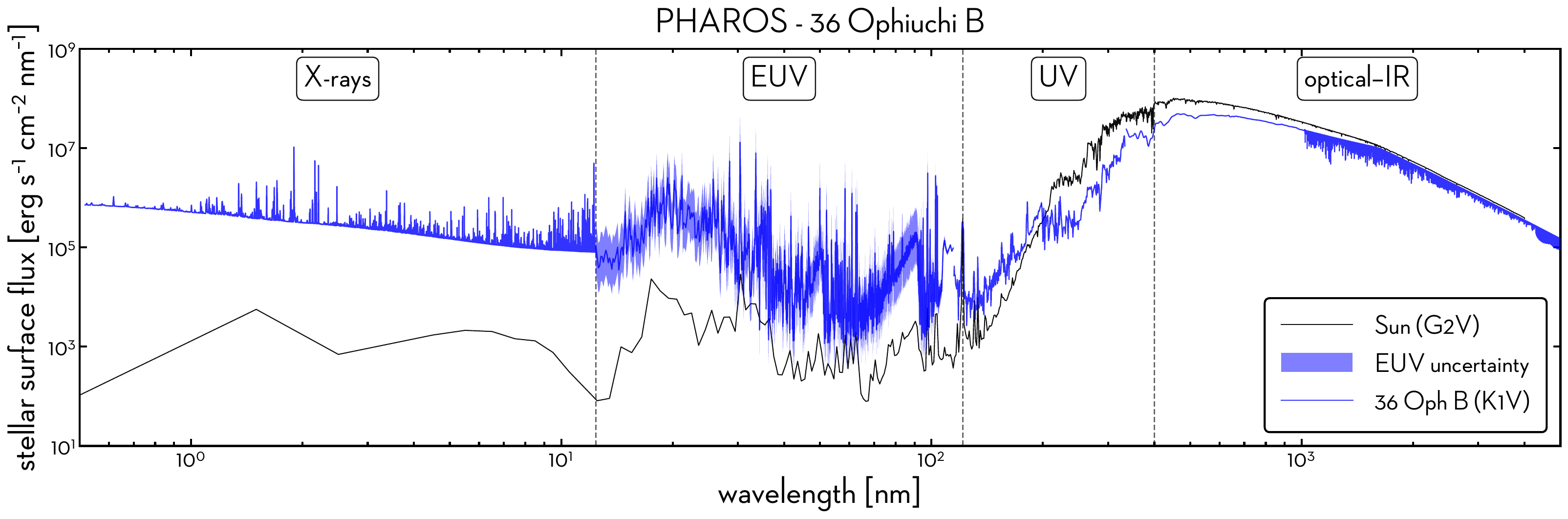}
    \includegraphics[width=0.87\textwidth]{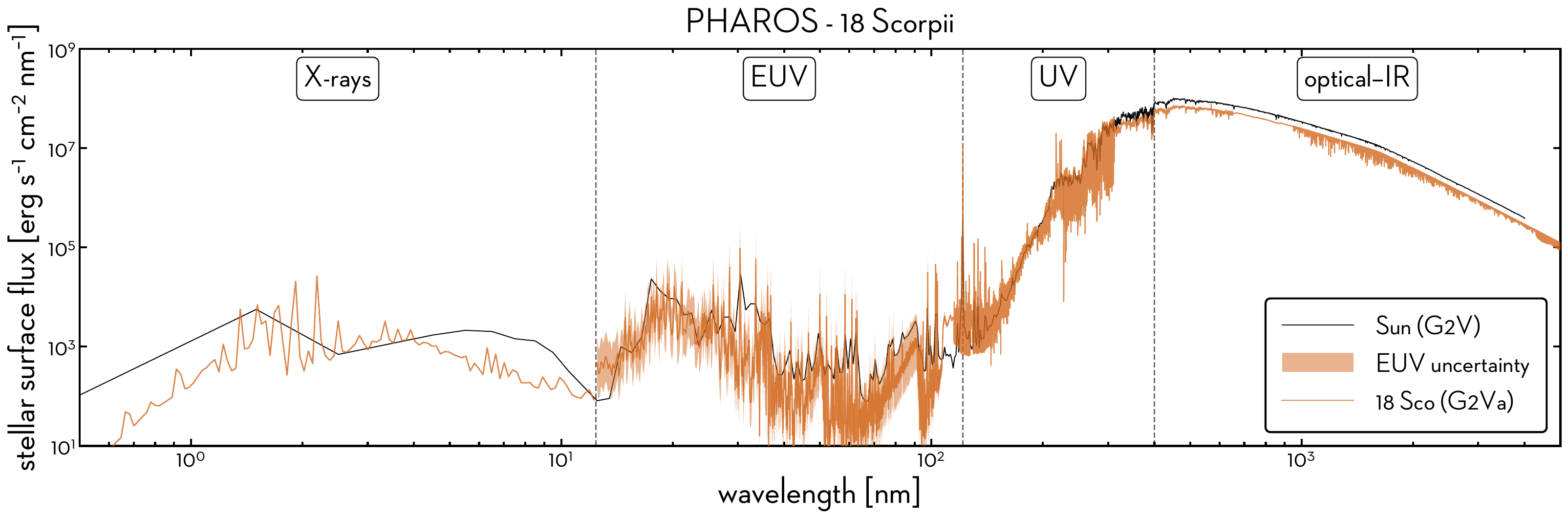}
    \includegraphics[width=0.87\textwidth]{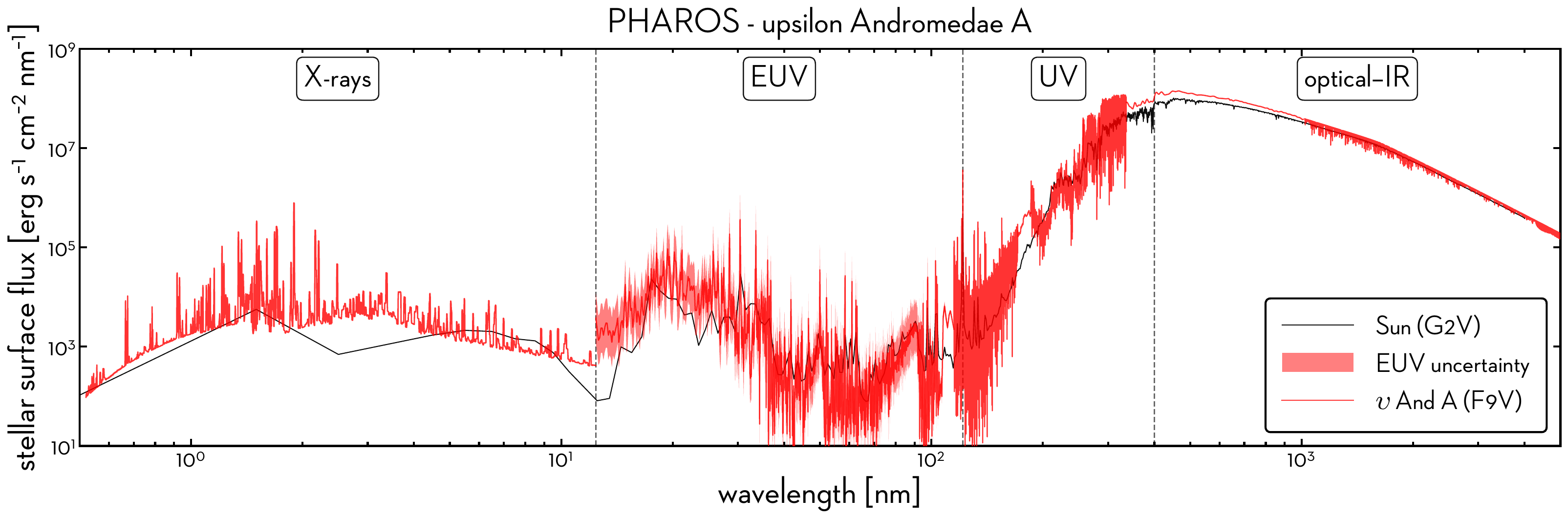}
    \caption{Final PHAROS stellar surface-flux SEDs for the Proterozoic-context group: 70~Oph~A, 36~Oph~B, 18~Sco, and $\upsilon$~And~A. The spectra are shown over 0.5--5000\,nm and compared with the reference solar spectrum \citep{Gueymard.2004SoEn...76..423G} using the same plotting convention as in Fig.~\ref{fig:pharos_seds_hadean}. The grouping is based on the adopted central stellar ages and is intended as a comparative age context rather than a direct evolutionary
    correspondence.}
    \label{fig:pharos_seds_proterozoic}
\end{figure*}

\begin{figure*}
\centering
    \includegraphics[width=0.87\textwidth]{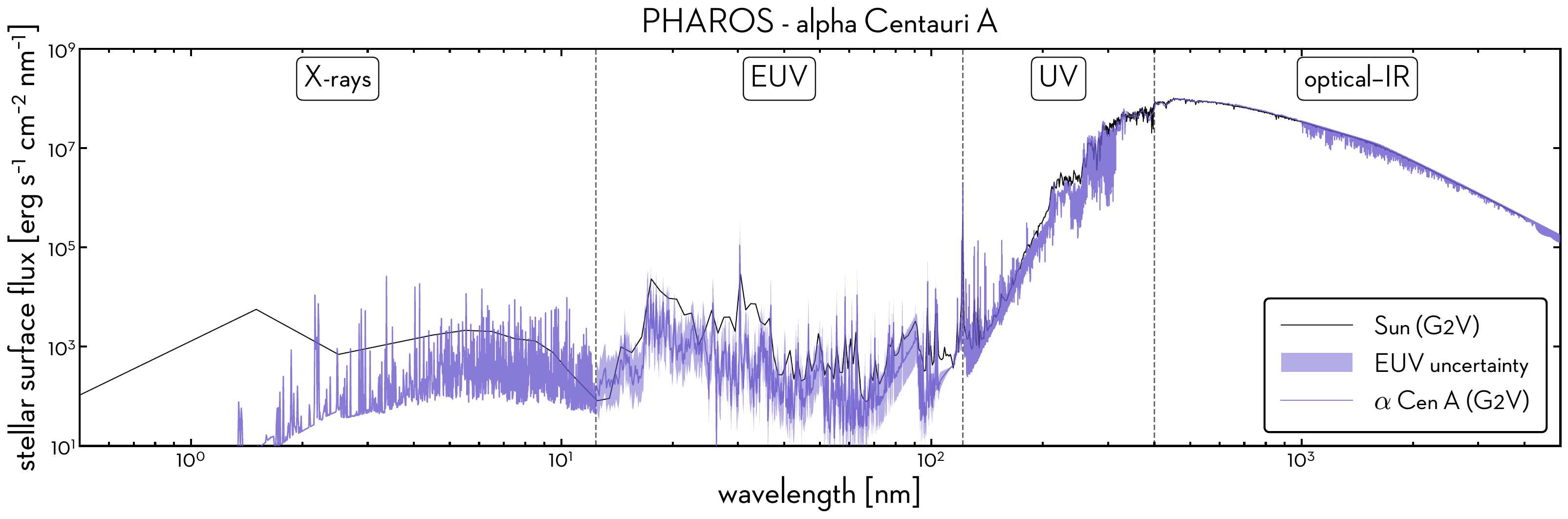}
    \includegraphics[width=0.87\textwidth]{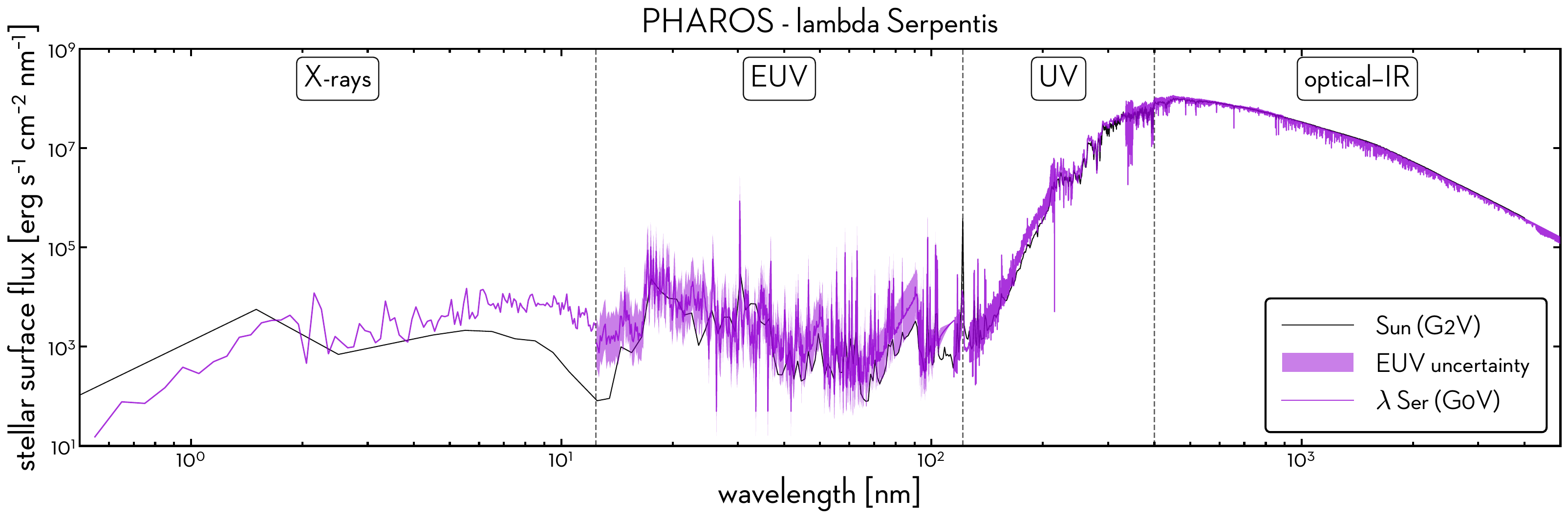}
    \includegraphics[width=0.87\textwidth]{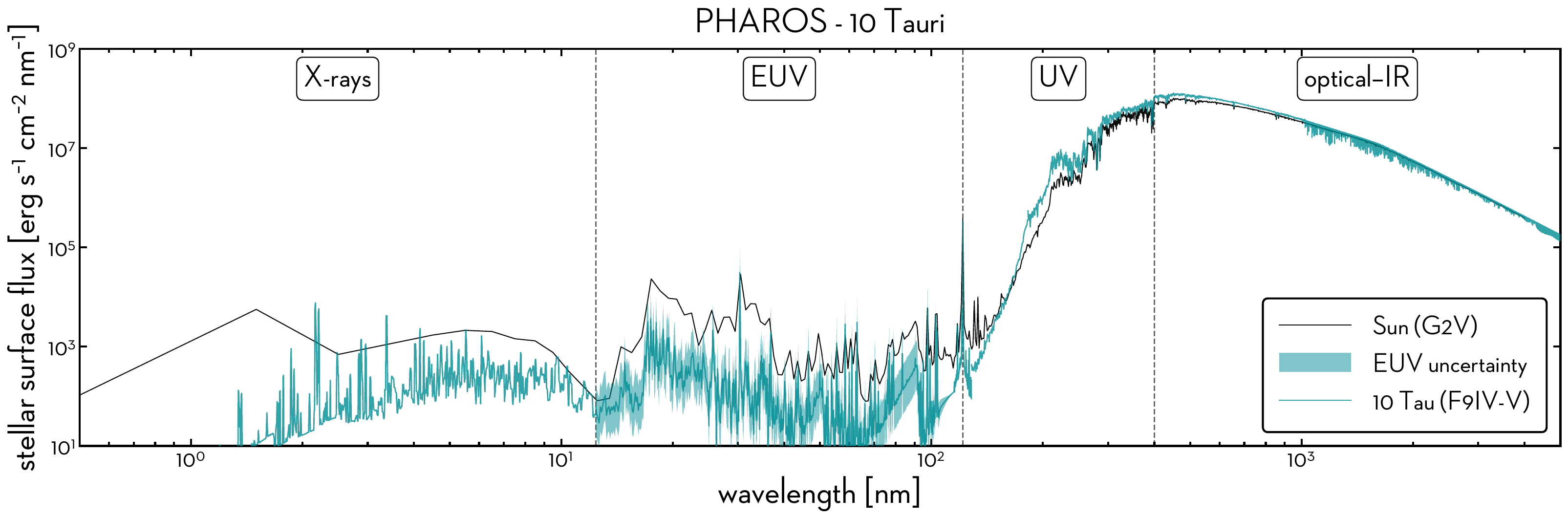}
    \includegraphics[width=0.87\textwidth]{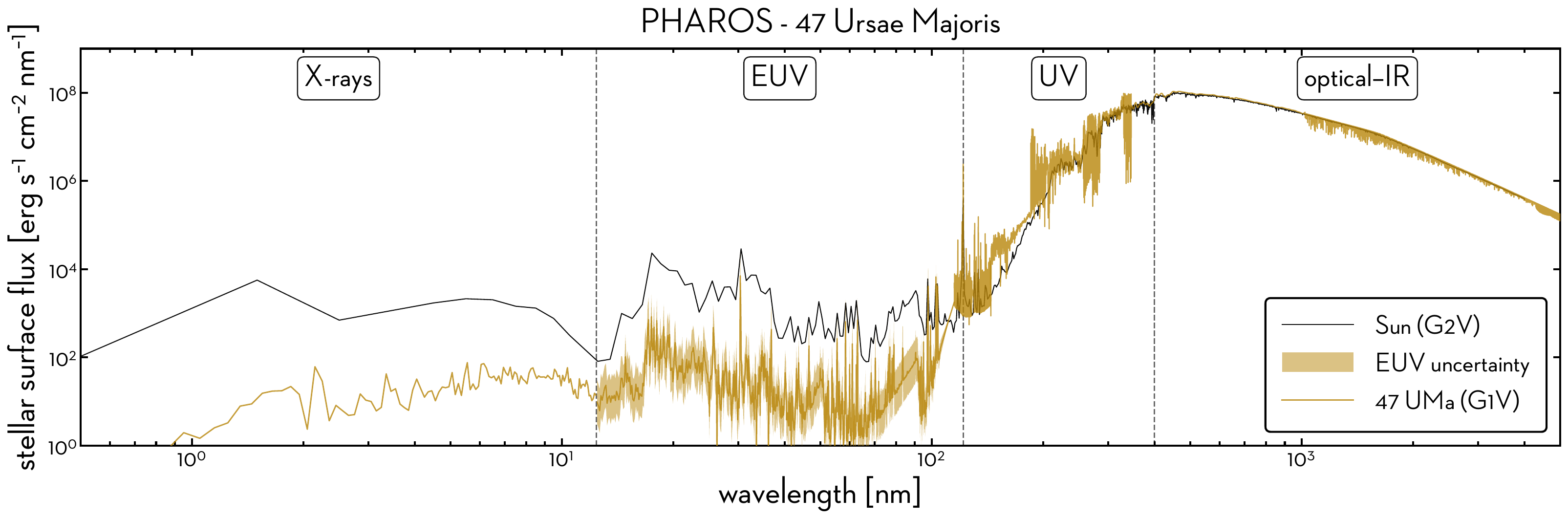}
    \caption{Final PHAROS stellar surface-flux SEDs for the older-than-Sun group: $\alpha$~Cen~A, $\lambda$~Ser, 10~Tau, and 47~UMa. The spectra are shown over 0.5--5000\,nm and compared with the reference solar spectrum \citep{Gueymard.2004SoEn...76..423G} using the same plotting convention as in Fig.~\ref{fig:pharos_seds_hadean}. These targets extend the comparison towards mature solar-like irradiation environments beyond the present solar age.}
    \label{fig:pharos_seds_older}
\end{figure*}

The first release delivers fifteen stellar surface-flux SEDs in a common 0.5--5000\,nm wavelength range. Each spectrum combines the adopted X-ray/EUV reconstruction, available ultraviolet and optical observations, and a long-wavelength PHOENIX extension.
The spectra are displayed in four age-ordered groups: Hadean context; Archean context, including targets near the Archean--Proterozoic transition; Proterozoic context; and older-than-Sun (Figs.~\ref{fig:pharos_seds_hadean}--\ref{fig:pharos_seds_older}).
The geological labels are used only as an organisational framework for
comparing stellar irradiation environments and do not imply that any individual target is an exact evolutionary analogue of the Sun during the corresponding interval. Stars whose age uncertainties cross a geological boundary are assigned to the group that provides the clearest comparative and graphical organisation, with the boundary overlap stated explicitly in the corresponding figure caption.

The strongest diversity appears at short wavelengths, where coronal and transition-region emission respond to rotation, magnetic activity, multiplicity, and evolutionary state. As a broad trend, the younger Hadean- and Archean/transition-context groups show enhanced X-ray, EUV, and ultraviolet output relative to their photospheric continua, whereas the Proterozoic-context and older-than-Sun groups generally occupy weaker high-energy regimes. The dispersion within each group, however, emphasises that the geological-era labels are age-context bins rather than deterministic predictors of stellar activity. These differences are particularly important for photochemical and atmospheric-escape calculations because the short-wavelength irradiation controls photolysis, ionisation, and upper-atmosphere heating.

At optical and infrared wavelengths, the spectra are dominated by the stellar photosphere and show smoother variations across the sample. In these regions, the differences among targets are driven primarily by effective temperature, stellar radius, and the scaling of the photospheric model extension to the observed optical continuum.

\begin{figure*}
\centering
    \includegraphics[width=0.67\columnwidth]{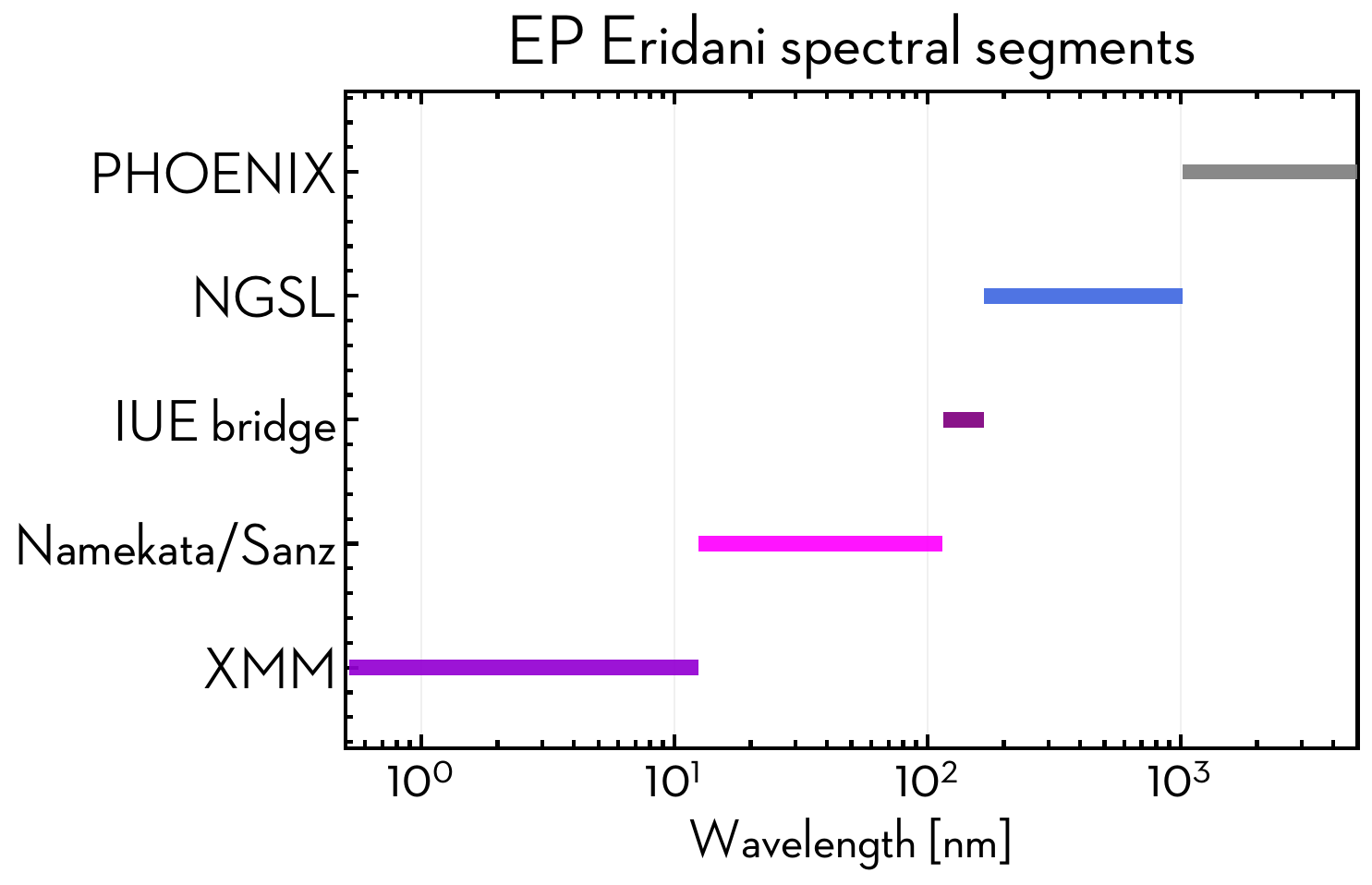}
    \includegraphics[width=0.67\columnwidth]{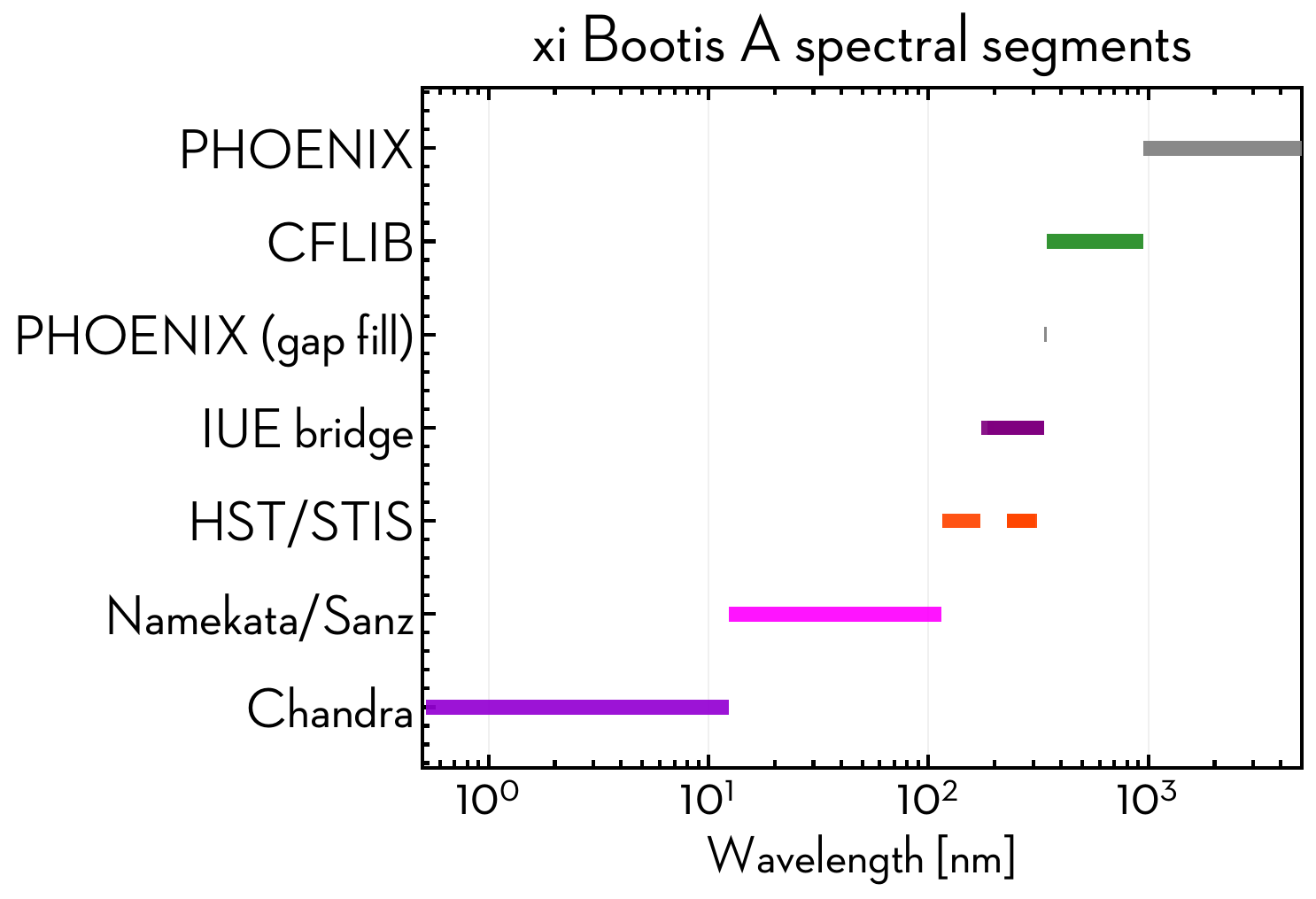}
    \includegraphics[width=0.67\columnwidth]{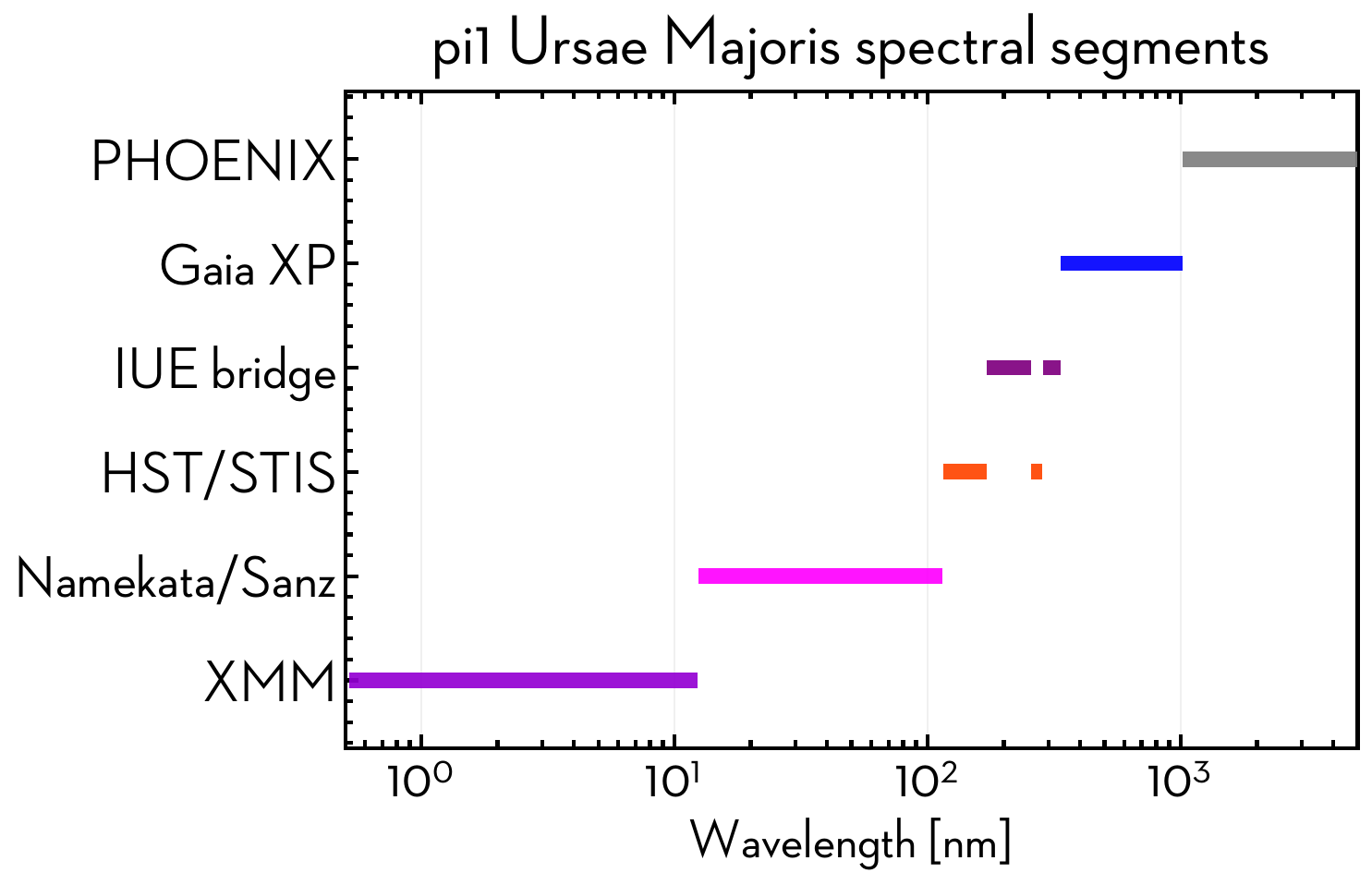}
    \includegraphics[width=0.67\columnwidth]{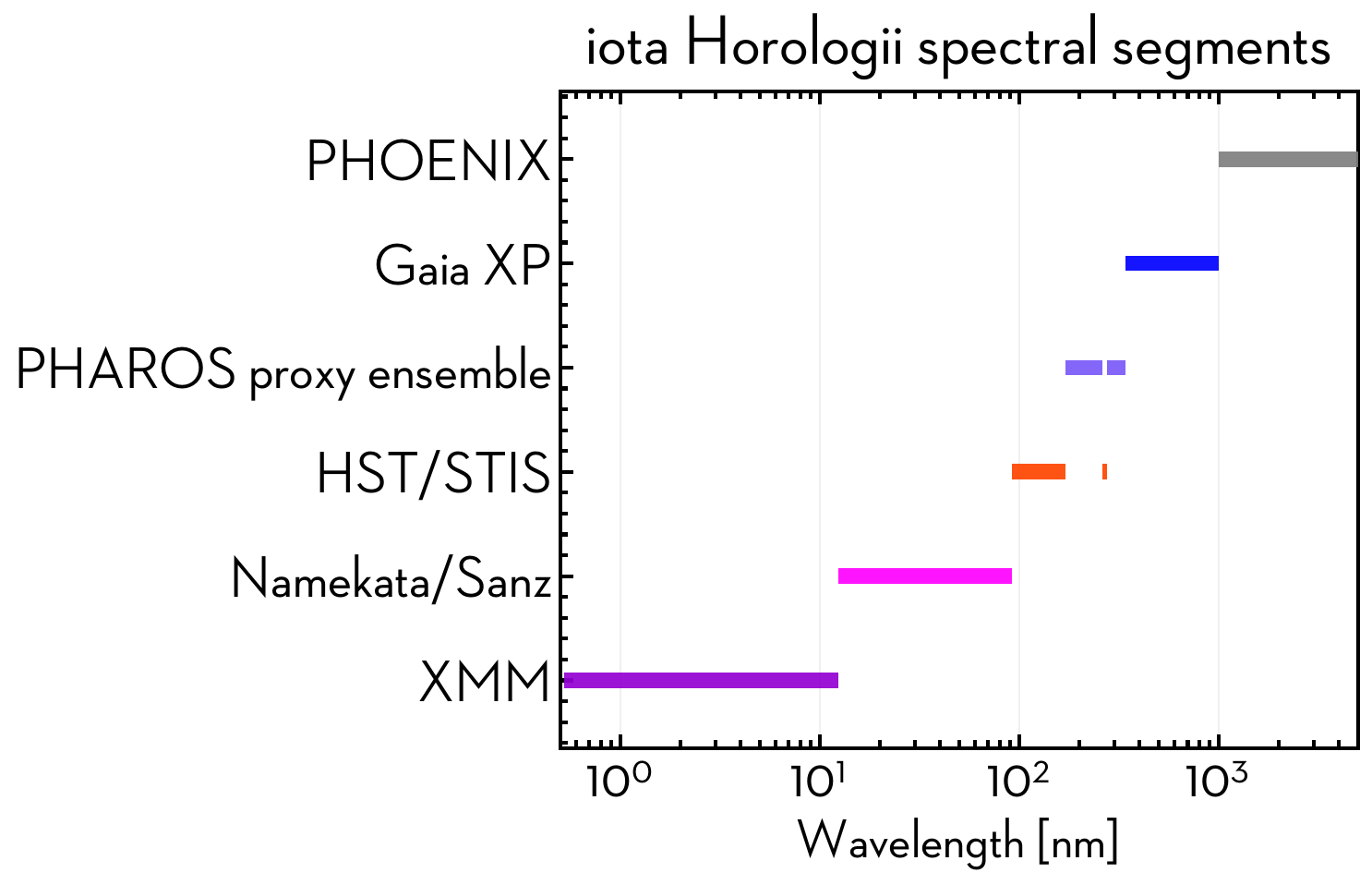}   
    \includegraphics[width=0.67\columnwidth]{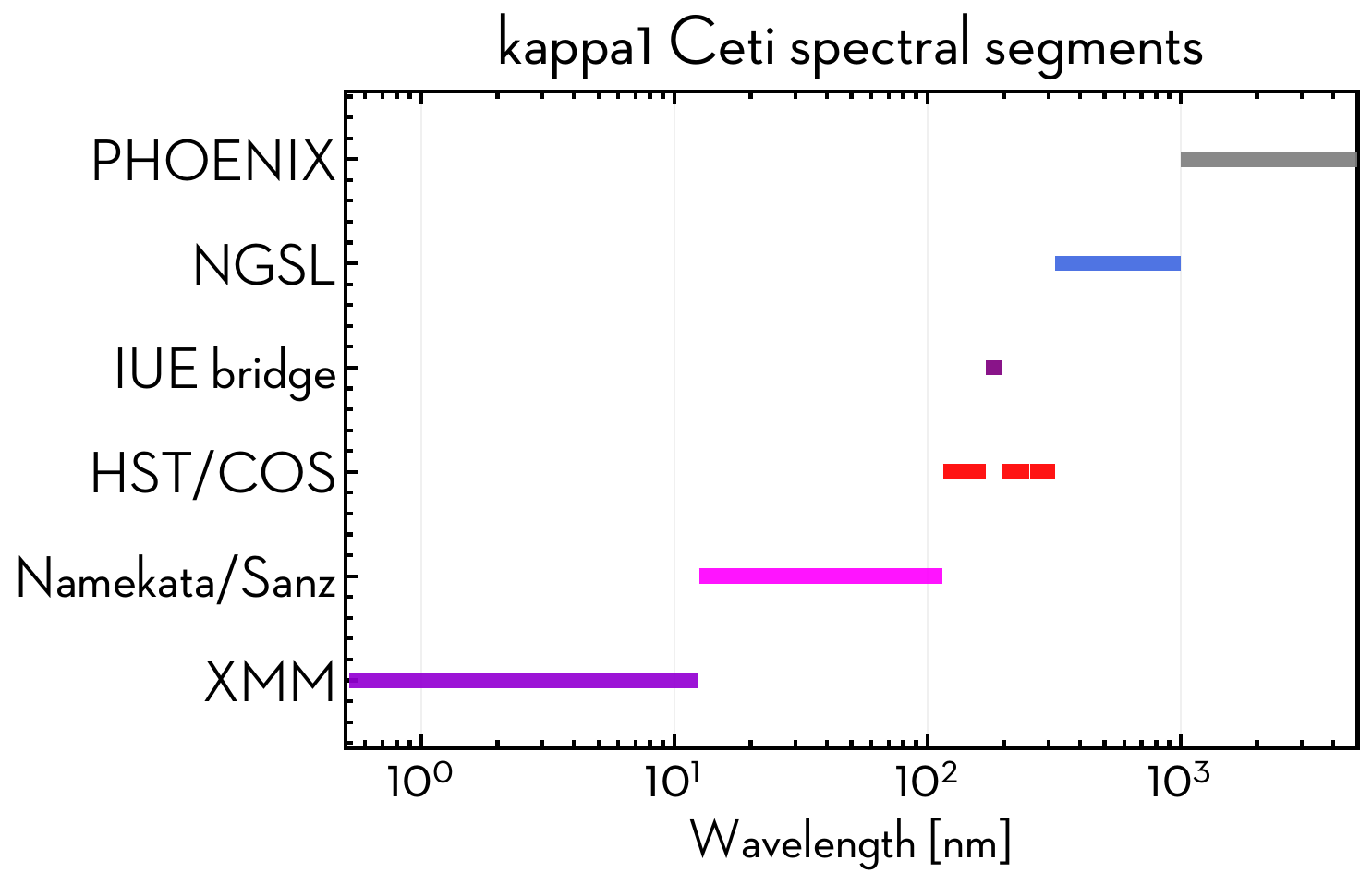}
    \includegraphics[width=0.67\columnwidth]{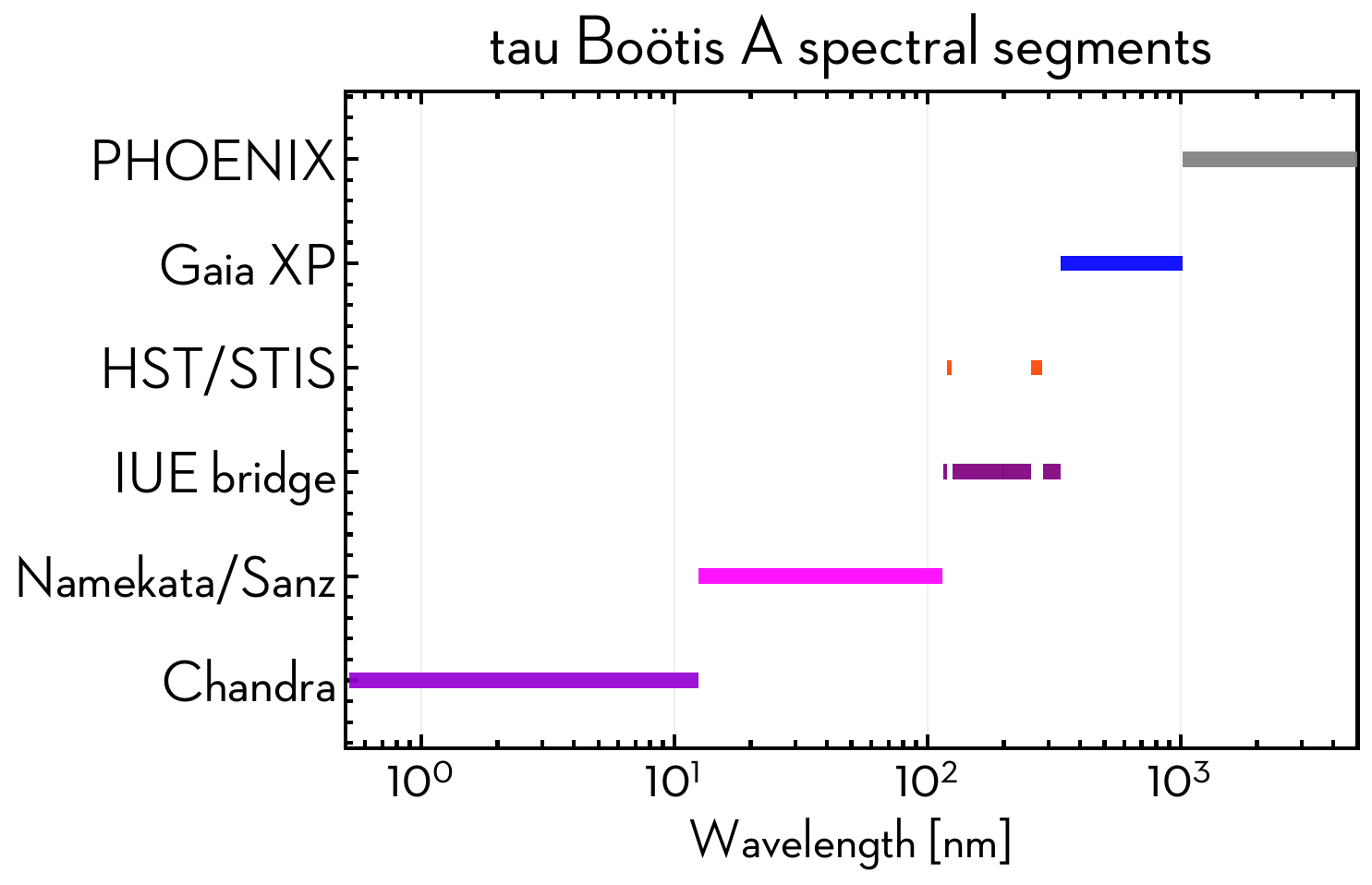}
    \includegraphics[width=0.67\columnwidth]{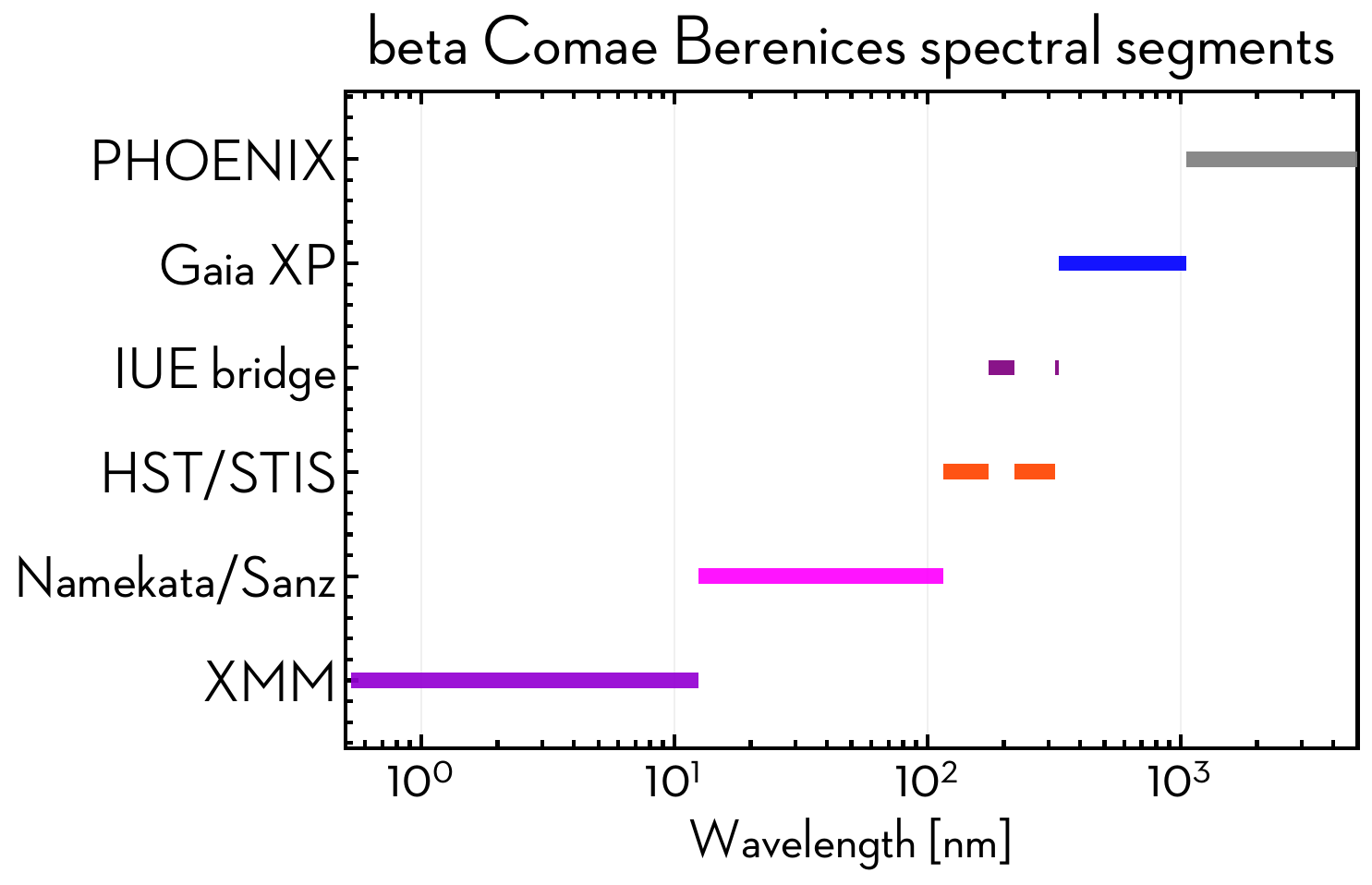}
    \includegraphics[width=0.67\columnwidth]{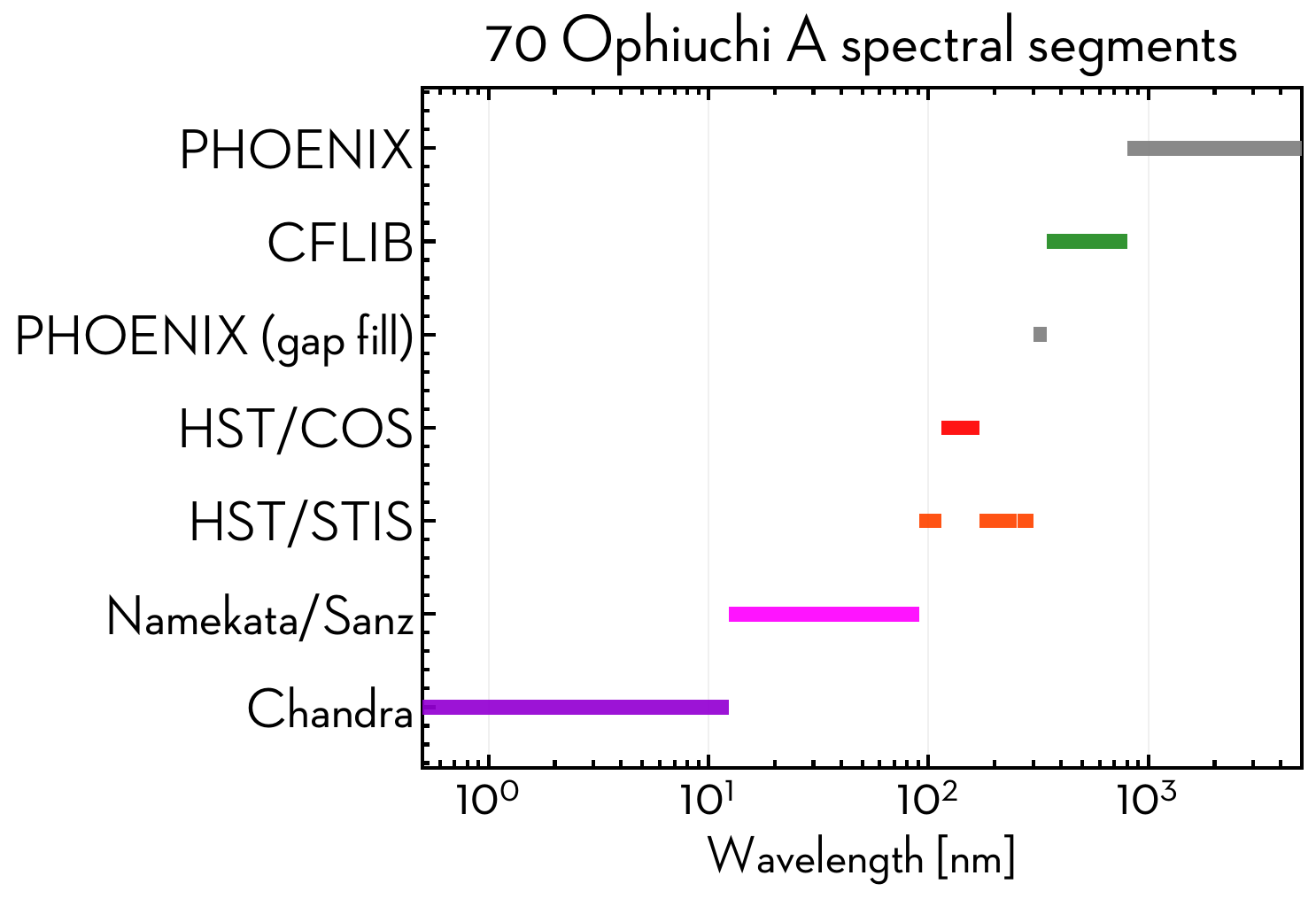}
    \includegraphics[width=0.67\columnwidth]{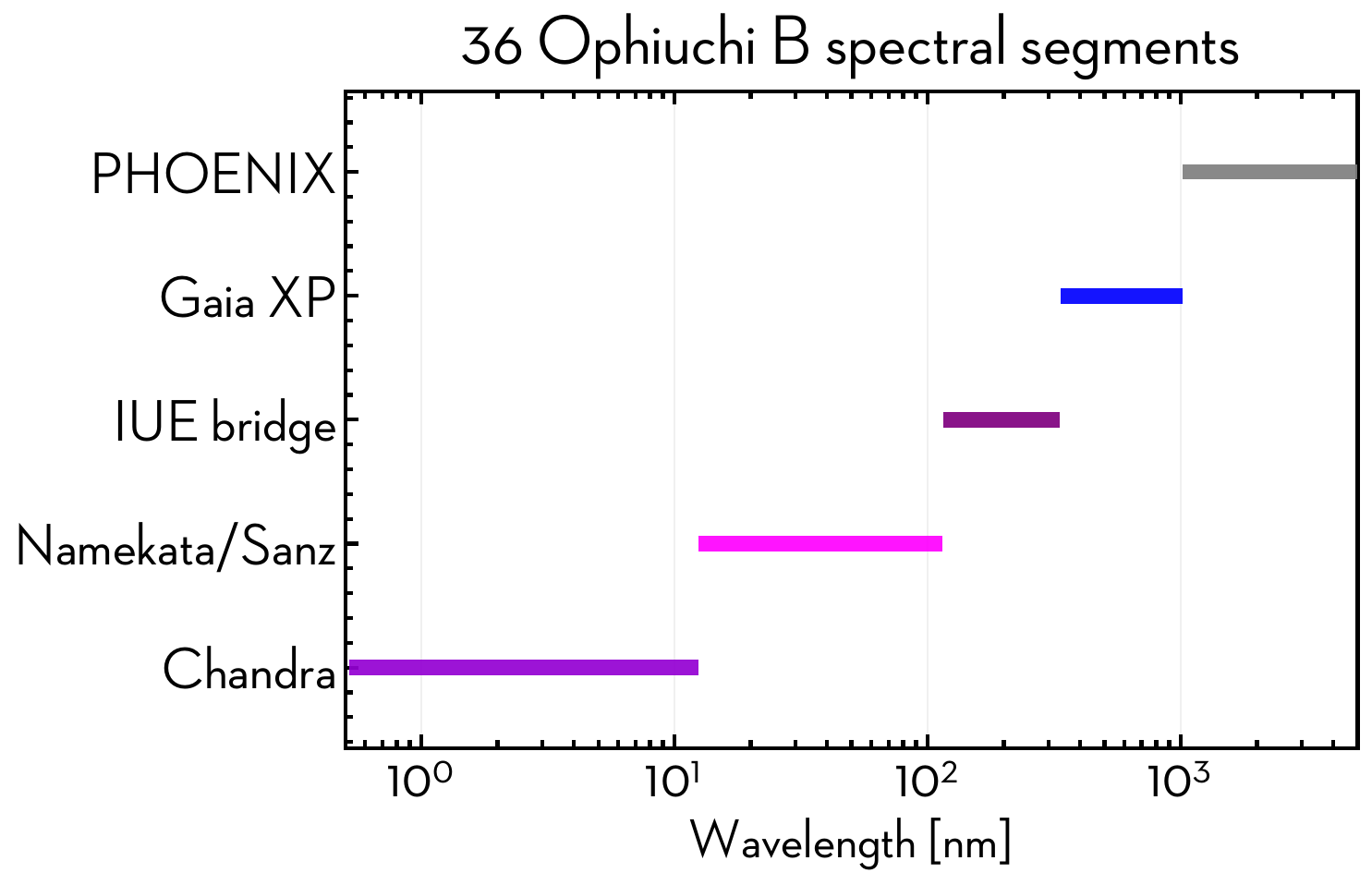}
    \includegraphics[width=0.67\columnwidth]{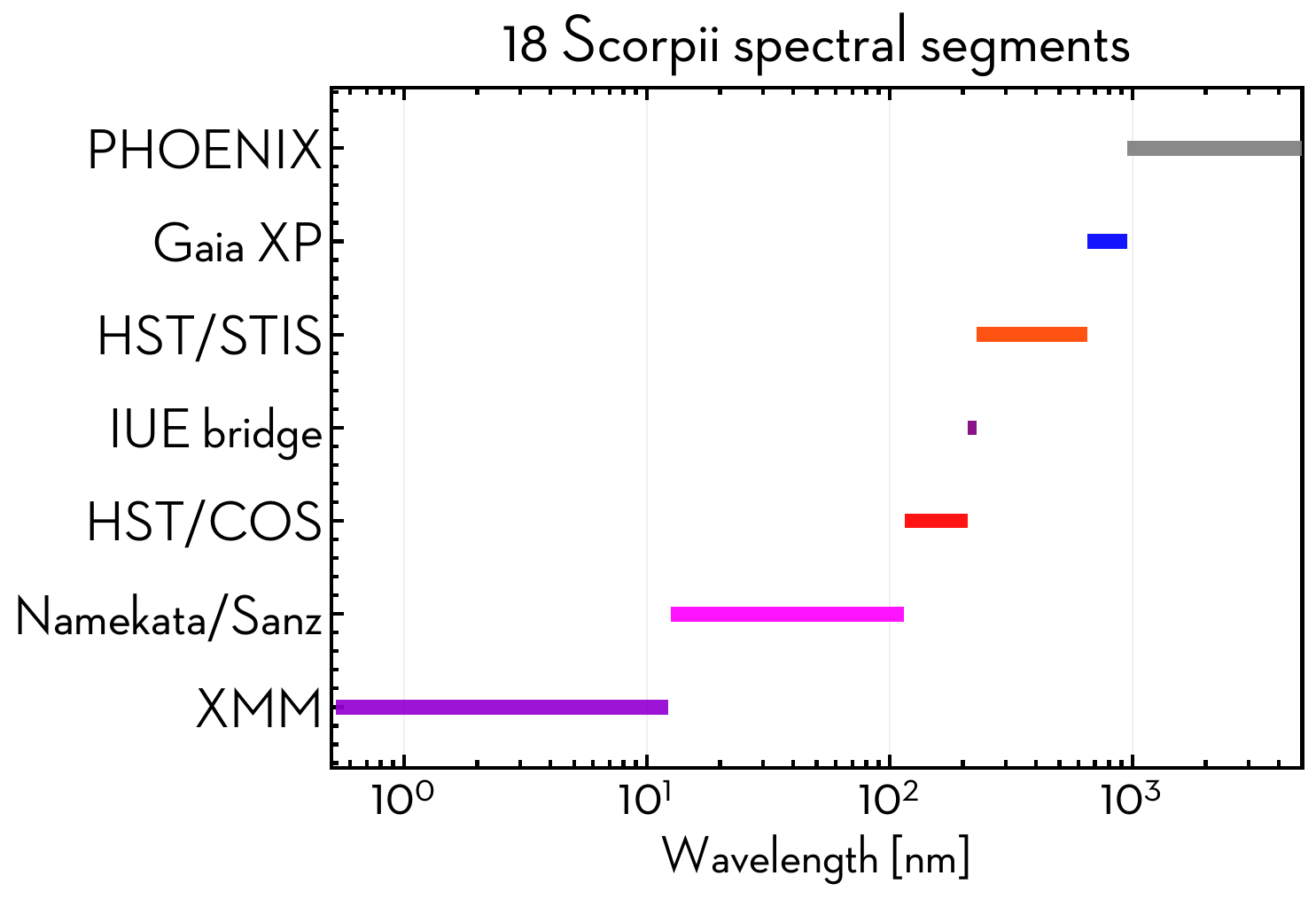}
    \includegraphics[width=0.67\columnwidth]{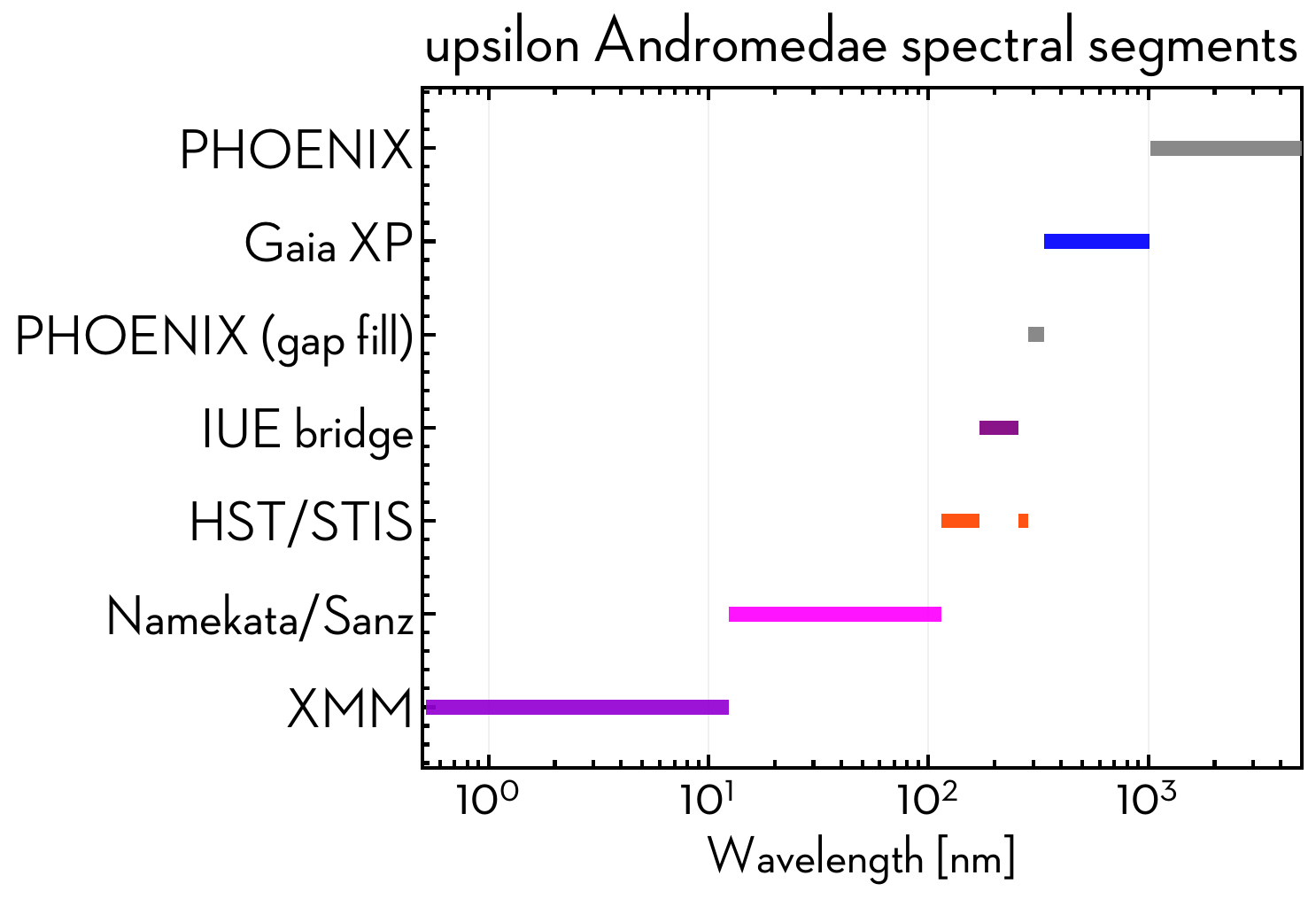}
    \includegraphics[width=0.67\columnwidth]{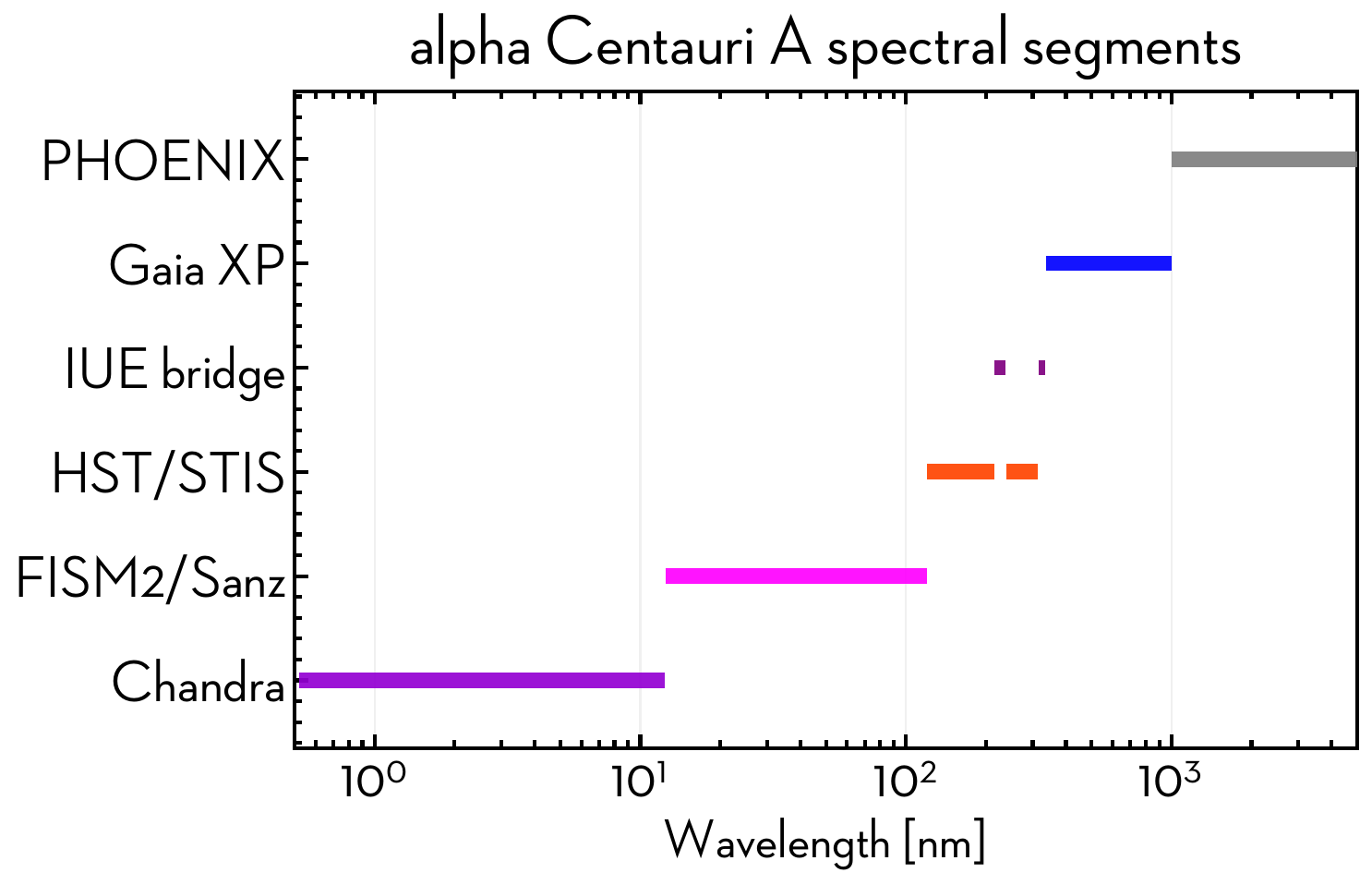}
    \includegraphics[width=0.67\columnwidth]{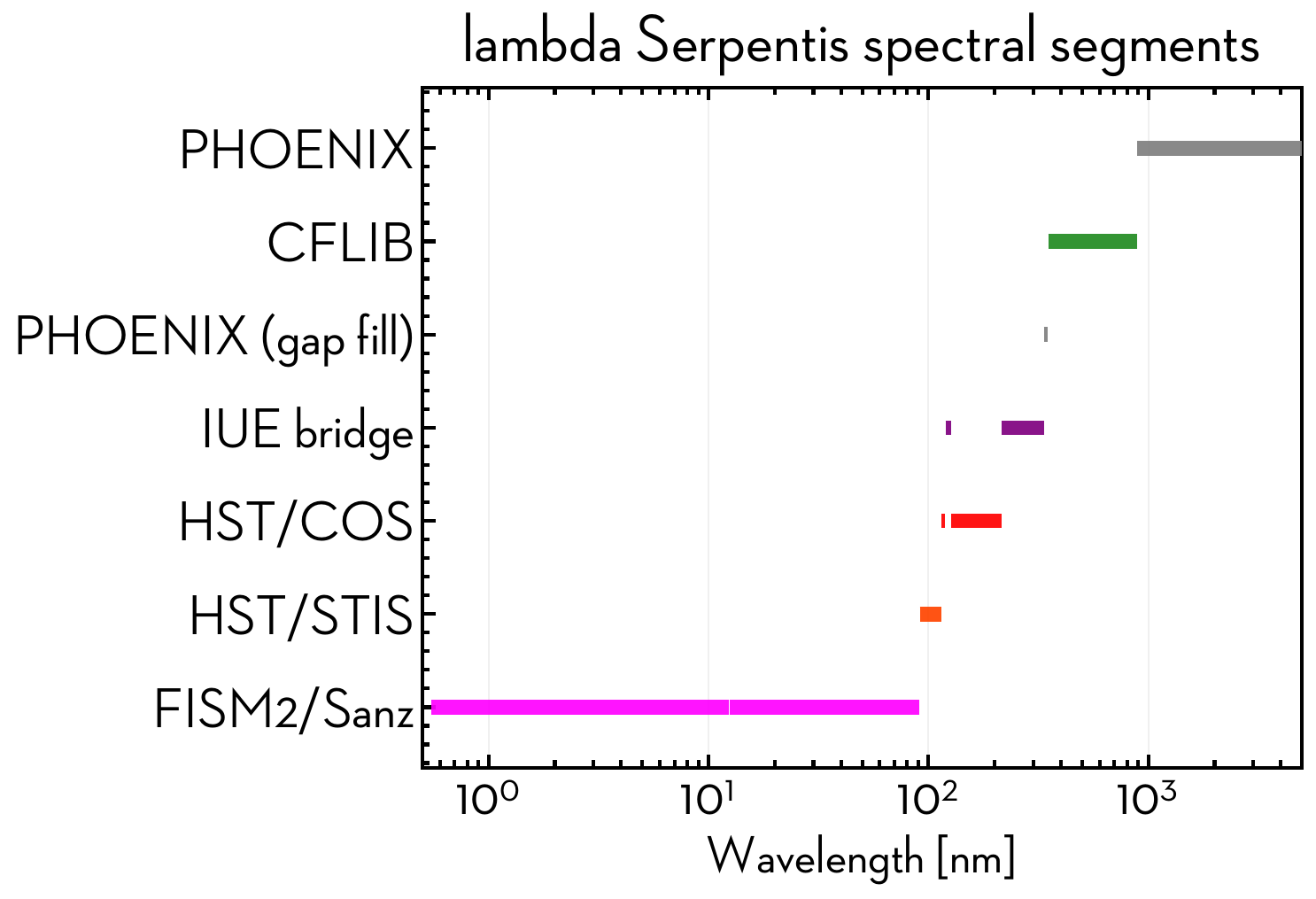}
    \includegraphics[width=0.67\columnwidth]{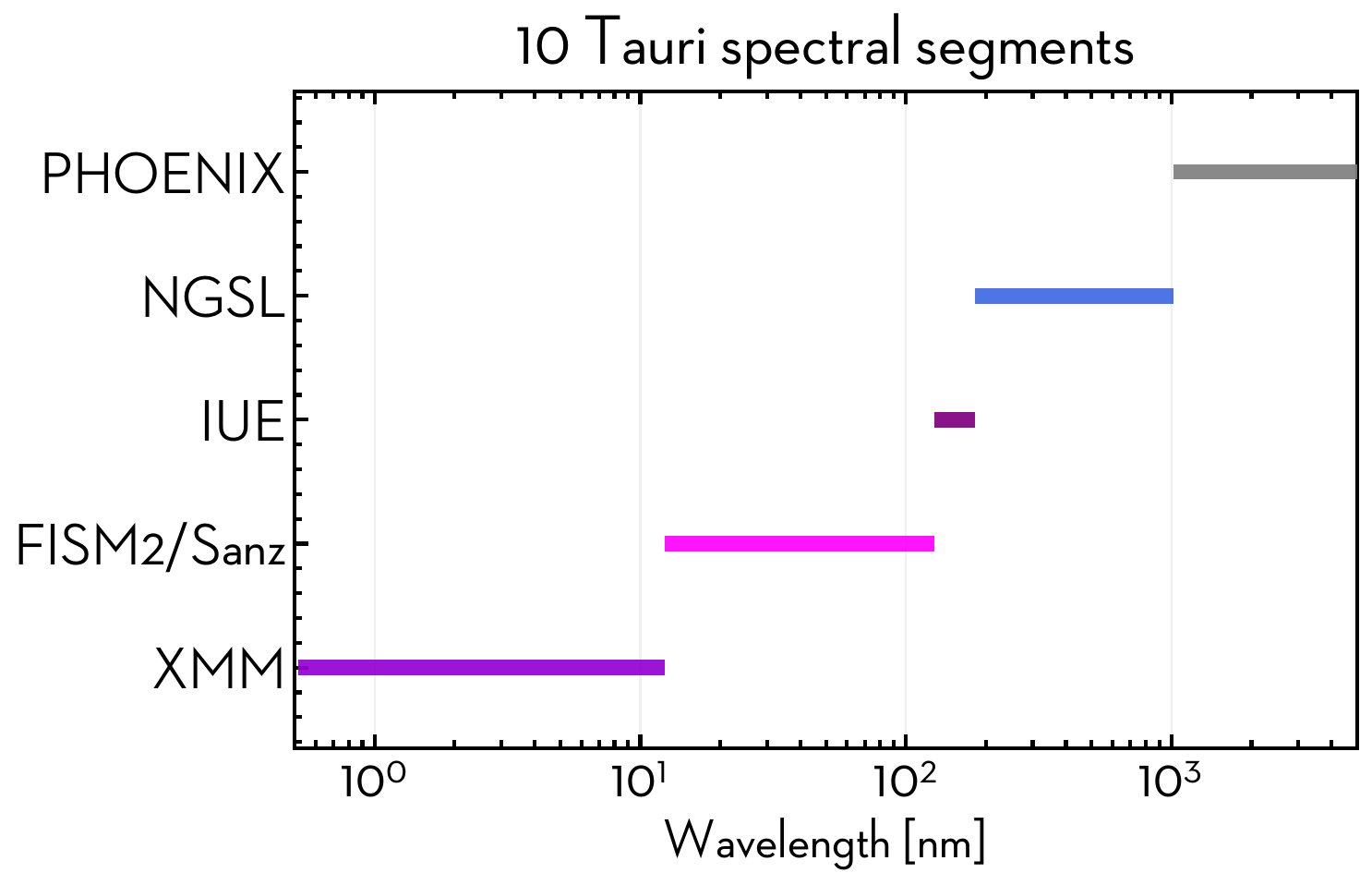}
    \includegraphics[width=0.67\columnwidth]{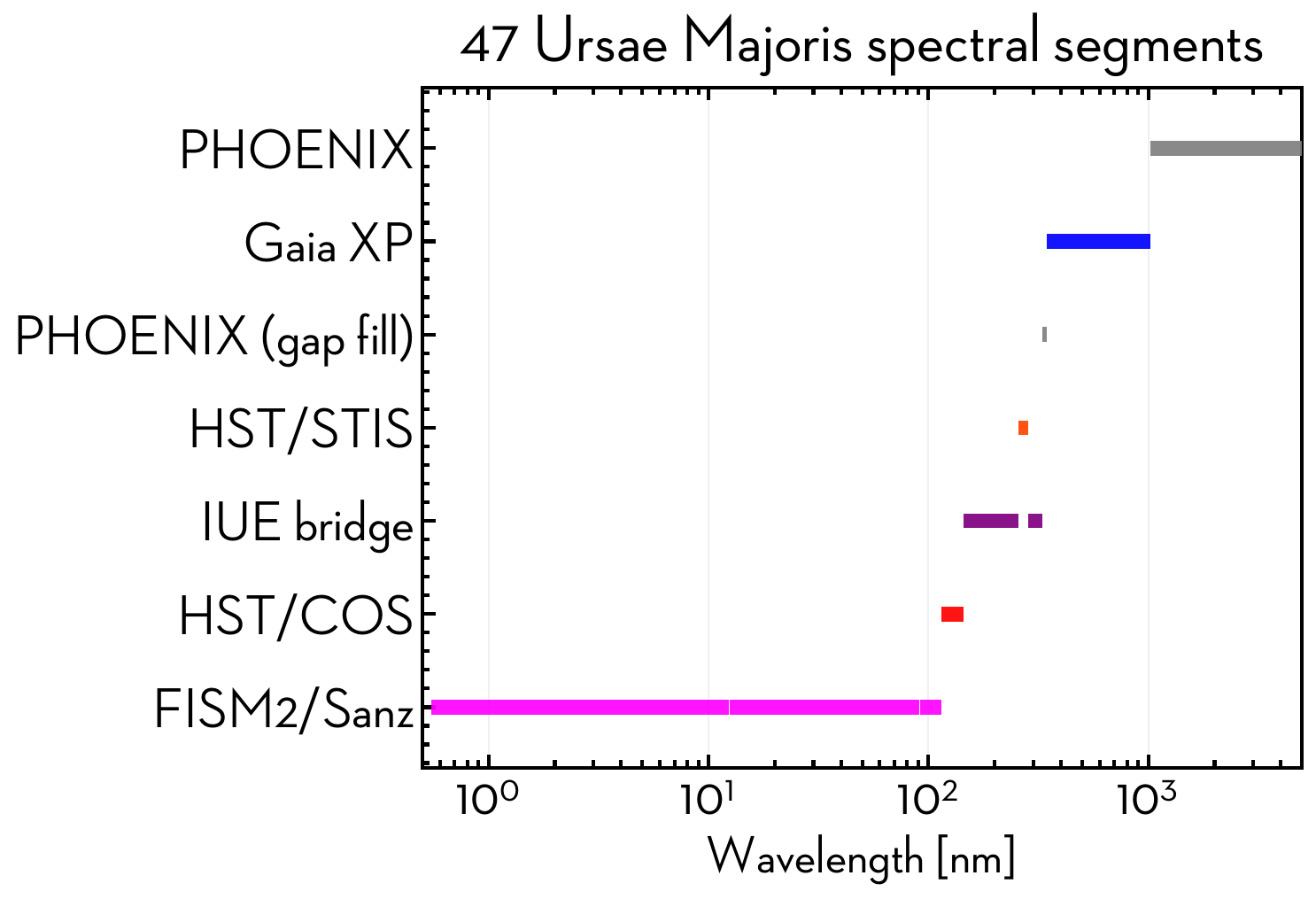}
    \caption{Component provenance maps for all fifteen targets in the first PHAROS release. Each panel identifies the wavelength intervals assigned to observed spectra, reconstructed regions, empirical templates, bridge segments, and photospheric models. The maps show that all products cover a comparable 0.5--5000\,nm range, while the observational or model origin of each interval remains target-dependent.}
    \label{fig:pharos_component_maps}
\end{figure*}

The component provenance of each stitched spectrum is summarised in Fig.~\ref{fig:pharos_component_maps}. These component maps identify which wavelength intervals are observationally constrained, reconstructed, empirically templated, or model-dependent. Although all products are delivered over a comparable 0.5--5000\,nm range, the supporting data differ substantially across the sample. Most targets use target-specific X-ray plasma models together with a
Namekata/Sanz EUV reconstruction. For $\alpha$~Cen~A and 10~Tau, target-specific APEC models are retained in the X-ray region while FISM2/Sanz provides the reconstructed EUV spectral shape. In contrast, $\lambda$~Ser and 47~UMa use FISM2-based empirical templates in both the soft-X-ray and EUV regions, normalised to the adopted luminosity constraints.

The UV and optical coverage also varies from target to target, with different combinations of \textit{HST}, \textit{IUE}, Gaia, NGSL, CFLIB, and PHOENIX components. The catalogue should therefore be used as a continuous irradiation input together with its component labels, which identify the observational or model basis of each spectral region.

\subsection{Target-specific reconstruction results}
\label{subsec:target_results}

Because the PHAROS SEDs combine heterogeneous observational, reconstructed, empirical-template, and model-dependent components, the final component sequence and the strength of the high-energy constraint differ from target to target. The following target-specific reconstruction results document the adopted X-ray anchor, EUV reconstruction, UV/optical components, final stitching choices, and the principal caveats associated with each stellar SED.

\subsubsection{EP~Eridani}

EP~Eridani (HD~17925) is the youngest K-type star in the PHAROS sample and was reconstructed as an active stellar irradiation field. The target is flagged as a possible close-binary case in the debris-disc literature \citep[e.g.,][]{Pearce.et.al.2022AA...659A.135P}, but its binary nature is not uniformly established.

The X-ray anchor is based on the \textit{XMM-Newton}/EPIC-MOS1 spectrum from ObsID~0203060501. The spectrum was fitted over 0.2--2.4\,keV in XSPEC/PyXspec using absorbed APEC plasma models, with the absorbing column fixed at $N_{\rm H}=10^{-4}\times10^{22}$\,cm$^{-2}$, appropriate for a nearby source. A single-temperature model was not adequate for the observed coronal spectrum, giving $C/{\rm dof}=2912.31/57$. We therefore adopted a two-temperature model, \texttt{tbabs*(apec+apec)}, which improved the fit to $C/{\rm dof}=353.66/55$. The fitted components have $kT_1=0.315$\,keV and $kT_2=0.885$\,keV, with APEC normalisations of $2.21\times10^{-3}$ and $1.21\times10^{-3}$, respectively. The corresponding unabsorbed luminosity is $L_X=7.86\times10^{28}$~erg~s$^{-1}$ over 0.2--2.0\,keV and $7.93\times10^{28}$~erg~s$^{-1}$ over 0.2--2.4\,keV. The MOS1 500\,s light curve does not show an obvious strong flare, so this model is used as a full-observation, time-averaged X-ray anchor.

The XMM/APEC model was exported as the stellar-surface X-ray segment from 0.52--12.4\,nm and used to normalise the semiempirical EUV reconstruction. Using the Sanz-Forcada relation, the adopted X-ray luminosity implies $L_{\rm EUV}=4.47\times10^{29}$~erg~s$^{-1}$ over 100--920\,\AA. The Namekata-based XUV/EUV spectral shape was then scaled so that the final integrated Namekata/Sanz luminosity ratio is unity. The stitched SED uses the Namekata/Sanz reconstruction from 12.4--115\,nm, IUE/SWP data from 115--167.6\,nm, NGSL/STIS from 167.6--1019.7\,nm, and a PHOENIX photospheric extension beyond 1019.7\,nm. The IUE/SWP segment is used only as a short-wavelength bridge to the NGSL spectrum and was scaled to the NGSL overlap, while the PHOENIX model was scaled to the red end of the NGSL continuum. The resulting EP~Eri spectrum can be used as a PHAROS irradiation field, provided that the high-energy output is interpreted as system-integrated if the proposed close-binary nature is real.

\subsubsection{$\xi$~Bo\"otis~A}

$\xi$~Bo\"otis~A (HD~131156~A; GJ~566~A) is a nearby active solar-like star in a visual binary system. For PHAROS, the high-energy spectrum was treated as a component-centred product for the A component. In Chandra ObsID~8899, the LETG/HRC-S zero-order image resolves the brighter A source and the fainter B source at an angular separation of $\simeq6.2$'', and the PHA2 extraction centroid is coincident with the A component to within $\simeq0.12$''. We therefore used the combined $\pm1$ first-order LETG/HRC-S spectrum as the X-ray anchor for $\xi$~Boo~A, with the interpretation caveat that any residual contribution from the companion is expected mainly at the longest dispersed wavelengths. The XSPEC/PyXspec fit was performed over 0.2--2.0\,keV using the C-statistic and a fixed absorbing column of $N_{\rm H}=10^{-4}\times10^{22}$\,cm$^{-2}$. A one-temperature absorbed APEC model gave $C/{\rm dof}=7125.46/4460$, while the adopted two-temperature model, \texttt{tbabs*(apec+apec)}, improved the fit to $C/{\rm dof}=6038.21/4458$. The fitted temperatures are $kT_1=0.313$\,keV and $kT_2=0.702$\,keV, with APEC normalisations of $1.16\times10^{-2}$ and $8.47\times10^{-3}$, respectively; the two abundance parameters were tied at $Z=0.300$ in the adopted XSPEC abundance scale. The corresponding unabsorbed luminosities are $L_X=8.77\times10^{28}$~erg~s$^{-1}$ over 0.2--2.0\,keV and $1.06\times10^{29}$~erg~s$^{-1}$ over 0.1--2.4\,keV.

The Chandra/APEC model was exported as the stellar-surface X-ray segment over 0.520--12.400\,nm and used to set the high-energy normalisation. For consistency with the Sanz-Forcada relation, the adopted 0.1--2.4\,keV luminosity implies $L_{\rm EUV}=5.78\times10^{29}$~erg~s$^{-1}$ over 100--920\,\AA. The Namekata-based spectral shape was then scaled by a factor of 14.34, yielding a final Namekata/Sanz luminosity ratio of unity. The stitched SED uses the Namekata/Sanz reconstruction from 12.4--91.19\,nm and a scaled Namekata extension from 91.21--114.5\,nm. The observed UV/NUV connection is built from HST/STIS E140M from 115.05--172.94\,nm, IUE/SWP over 173.06--197.87\,nm, and a priority-merged combination of IUE/LWP, IUE/LWR, and HST/STIS E230M products through the NUV up to $\simeq334.95$\,nm, with HST preferred where overlapping data exist. Although archival EUVE products were inspected, they were retained only as qualitative diagnostics and were not used in the catalogue stitch.

The photospheric connection uses the nearest suitable PHOENIX grid spectrum, which was empirically scaled by a factor of 0.974 to the observed HST/IUE continuum over 320--335\,nm.
This PHOENIX spectrum provides a short bridge from 334.996--346.446\,nm, before the empirical Indo-US/CFLIB spectrum begins at 346.5\,nm. The CFLIB relative spectrum was calibrated to the empirically scaled PHOENIX continuum over 500--850\,nm, with a scale factor of $6.45\times10^{7}$, and was used from 346.5--946.9\,nm. The final red extension is the same scaled PHOENIX model from 946.95\,nm to 5000\,nm. The resulting $\xi$~Boo~A SED is therefore a usable component-centred PHAROS irradiation field, with the appropriate caveat that the high-energy extraction is centred on the resolved A component while the longest-wavelength dispersed response and the broader UV--optical calibration should be interpreted within the binary-system context.

\subsubsection{\texorpdfstring{$\pi^1$~Ursae~Majoris}{pi1 Ursae Majoris}}

$\pi^1$~Ursae~Majoris (HD~72905) is a young solar analogue and is treated in PHAROS as an active, single-star irradiation field, without the component-separation caveats required for visual binaries such as 70~Oph. The available \textit{XMM-Newton}/EPIC-pn PPS light curve is mostly stable at the 500~s level, with one elevated bin rather than a sustained flare interval. We therefore use the full-observation spectrum as a time-averaged, near-quiescent X-ray anchor while retaining the caveat that no event-level GTI filtering was applied.

The X-ray segment is based on the \textit{XMM-Newton}/EPIC-pn spectrum from ObsID~0111400101, source~0000, chain PNS003, fitted over 0.2--2.4~keV in XSPEC/PyXspec with C-statistics. The absorption column was fixed at $N_{\rm H}=10^{-4}\times10^{22}$\,cm$^{-2}$, which is appropriate for a nearby source. A single-temperature APEC model was inadequate, giving $C/{\rm dof}=1216.39/45$. We therefore adopted a two-temperature \texttt{tbabs*(apec+apec)} model with the APEC abundance tied between the two components and allowed to vary; this abundance-free model gave $C/{\rm dof}=150.37/42$, substantially better than the fixed-abundance alternatives. The fitted temperatures are $kT_1=0.418$\,keV and $kT_2=0.740$\,keV, with tied abundance $Z=0.448\,Z_\odot$ and XSPEC normalisations of $2.88\times10^{-3}$ and $8.84\times10^{-4}$, respectively. The corresponding unabsorbed luminosities are $L_X(0.2$--$2.0\,{\rm keV})=9.93\times10^{28}$\,erg~s$^{-1}$ and $L_X(0.2$--$2.4~{\rm keV})=1.00\times10^{29}$\,erg~s$^{-1}$; the exported 0.1--2.4\,keV surface-flux grid gives $L_X=1.16\times10^{29}$\,erg~s$^{-1}$.

The adopted Sanz-Forcada input is the unabsorbed 0.2--2.0\,keV luminosity, $L_X=9.93\times10^{28}$\,erg~s$^{-1}$, which gives $L_{\rm EUV}=5.46\times10^{29}$\,erg~s$^{-1}$ over 100--920\,\AA. The Namekata spectrum was scaled by a factor of 13.56 over 12.4--91.2\,nm, yielding a final Namekata/Sanz luminosity ratio of unity. The stitched SED uses the abundance-free XMM/APEC model from 0.518--12.37\,nm, the Namekata/Sanz reconstruction from 12.4--114.5\,nm, HST spectral products from 115.0--171.0\,nm and 256.9--284.5\,nm, IUE/SWP and IUE/LWR+LWP bridge segments over 171.0--256.7\,nm and 284.7--334.8\,nm, Gaia DR3 XP from 336--1020\,nm, and a PHOENIX photospheric extension from 1020\,nm to 5\,$\mu$m. Gaia XP is kept in its native absolute calibration, while the PHOENIX extension is scaled to the Gaia red optical anchor by a factor of 0.971.

\subsubsection{\texorpdfstring{$\iota$~Horologii}{iota Horologii}}

$\iota$~Horologii ($\iota$~Hor, HD~17051) is a young F-type planet-host star with an extensive \textit{XMM-Newton} archive. The available observations were screened before selecting the high-energy anchor, rather than simply adopting the exposure with the highest count rate. We compared the EPIC light curves and spectral products to identify an observation representative of a relatively quiet coronal state. The preferred data set is the EPIC-MOS1 M1S001 source~0001 spectrum from ObsID~0693550701, with an effective exposure of 6548.8~s and 1067 good counts. Its associated light curve shows no clear flare-like enhancement and was therefore preferred over the brighter EPIC-pn candidate from ObsID~0693550501, whose temporal diagnostics indicate an activity-enhanced interval. The adopted MOS1 spectrum is consequently treated as a time-averaged quiet-state X-ray anchor rather than as a characterisation of the full coronal variability of the star.

The spectrum was fitted over 0.2--2.4\,keV in XSPEC/PyXspec using absorbed APEC models and the C-statistic. We adopted \texttt{tbabs} absorption with $N_{\rm H}=10^{-4}\times10^{22}$\,cm$^{-2}$ fixed, Wilms abundances, Verner photoelectric cross sections, the APEC abundance fixed at the adopted solar value, and zero redshift. The single-temperature model, \texttt{tbabs*apec}, gives $C/{\rm dof}=22.63/24$, with $kT=0.439$\,keV and an APEC normalisation of $7.47\times10^{-4}$. A two-temperature model reduces the statistic to $C/{\rm dof}=19.25/22$, with $kT_1=0.419$\,keV and $kT_2=0.827$\,keV, but does not improve the information criterion: the AIC values are 26.63 and 27.25 for the 1T and 2T models, respectively. Moreover, the inferred 0.2--2.4\,keV luminosity changes by less than one percent between the two models. We therefore adopt the simpler 1T representation as the nominal coronal model. Its unabsorbed flux is $F_X(0.2$--$2.4\,{\rm keV})=1.44\times10^{-12}$\, erg\,cm$^{-2}$\,s$^{-1}$, corresponding to $L_X(0.2$--$2.4\,{\rm keV})=5.13\times10^{28}$\,erg\,s$^{-1}$ at the adopted distance of 17.24\,pc. Over 0.1--2.4\,keV, the corresponding luminosity is $5.74\times10^{28}$\,erg\,s$^{-1}$. The unabsorbed 1T APEC model was exported as a stellar-surface $F_\lambda$ spectrum over 0.521--12.372\,nm and used as the X-ray component of the PHAROS SED.

The adopted X-ray luminosity sets the normalisation of the unobserved EUV region through the Sanz-Forcada relation, giving $L_{\rm EUV}\simeq3.10\times10^{29}$\,erg\,s$^{-1}$ over 100--920\,\AA. The Namekata-based spectral shape was scaled so that its integrated luminosity over 12.4--91.2\,nm matches this value, followed by a smooth transition to the observed ultraviolet spectrum. The final SED uses the XMM/MOS1 1T APEC model from 0.521--12.372\,nm, Namekata/Sanz from 12.400--91.990\,nm, and the Namekata/Sanz-to-HST transition from 92.010--114.500\,nm. HST/STIS E140M provides the directly observed spectrum from 115.005--170.971\,nm. Because no suitable target-specific spectrum continuously covers the interval to the E230H observation, the 171.021--260.951\,nm gap is reconstructed from a weighted PHAROS ensemble of $\kappa^1$~Cet, $\pi^1$~UMa, and $\upsilon$~And~A, with each proxy independently anchored to the $\iota$~Hor continua at the two boundaries. HST/STIS E230H is retained over its cleaner 261.001--277.000\,nm interval after removal of the narrow artefacts at both spectral edges. A second endpoint-anchored proxy ensemble fills 277.050--339.950\,nm, followed by Gaia DR3 XP from 340--1000\,nm and a PHOENIX photospheric extension scaled to Gaia from 1000.077\,nm to 5\,$\mu$m. The Namekata/Sanz normalisation uncertainty is stored as the analytic log-normal envelope described in Sect.~\ref{subsec:euv_reconstruction}, while the proxy-to-proxy 16th--84th percentile spread is retained separately as a conditional uncertainty on the shapes of the two reconstructed UV intervals.

\subsubsection{\texorpdfstring{$\kappa^1$~Ceti}{kappa1 Ceti}}

$\kappa^1$~Ceti (HD~20630) is a young, magnetically active solar analogue, and its PHAROS SED is treated as a single-star irradiation field. Several \textit{XMM-Newton} EPIC-pn observations were inspected, and ObsID~0822791001 was adopted as the X-ray anchor because it provides the best-ranked 2T fit among the tested PN source~0000 spectra while remaining in a quiet-to-moderate light-curve state. In the 300~s rebinned light curve, the adopted observation has $p_{95}/{\rm baseline}=1.14$ and ${\rm max}/{\rm baseline}=1.27$, with no bins flagged as a possible flare. We therefore use the full-observation spectrum as a time-averaged active-state X-ray anchor.

The X-ray spectrum was fitted in XSPEC/PyXspec over 0.2--2.4\,keV using C-statistics and an absorbed APEC model with $N_{\rm H}=10^{-4}\times10^{22}$\,cm$^{-2}$ fixed. For the adopted ObsID~0822791001 EPIC-pn source~0000 spectrum, a single-temperature model is inadequate, giving $C/{\rm dof}=7927.38/46$. We therefore adopted a two-temperature model, \texttt{tbabs*(apec+apec)}, with solar abundance fixed at $Z=1$ and zero redshift. The fitted components have $kT_1=0.709$\,keV and $kT_2=0.340$~keV, with XSPEC normalisations of $5.12\times10^{-4}$ and $2.28\times10^{-3}$, respectively, and give $C/{\rm dof}=94.70/44$. The corresponding unabsorbed luminosities are $L_X(0.2$--$2.0\,{\rm keV})=5.25\times10^{28}$\,erg~s$^{-1}$ and $L_X(0.2$--$2.4\,{\rm keV})=5.28\times10^{28}$\,erg~s$^{-1}$.

The adopted XMM/APEC model was exported as the stellar-surface X-ray segment from 0.520--12.394\,nm. The Sanz-Forcada normalisation uses $L_X(0.2$--$2.0~{\rm keV})=5.26\times10^{28}$\,erg~s$^{-1}$, giving $L_{\rm EUV}=3.16\times10^{29}$\,erg~s$^{-1}$ over 12.4--91.2\,nm. The Namekata spectral shape was scaled so that its integrated EUV luminosity matches this value, reducing the unscaled Namekata/Sanz ratio from 4.87 to unity; the adopted EUV total scale is $5.68\times10^{-2}$, with an HST/UV-overlap scale of 0.277 and an overlap scatter of 0.156\,dex. The final stitch uses XMM/EPIC-pn 2T APEC from 0.520--12.394\,nm, Namekata/Sanz EUV from 12.5--91.19\,nm, a Namekata/Sanz-to-UV transition from 91.21--114.5\,nm, HST/COS from 115.01--169.87\,nm, IUE/SWP as a short bridge from 170.04--197.37\,nm, HST/STIS E230 from 197.5--251.06\,nm and 253.52--319.74~nm, HST/STIS NGSL from 320.02--1000.0\,nm, and a PHOENIX photospheric extension from 1000.08\,nm to 5\,$\mu$m. The noisy IUE/LWP+LWR segment was not used, and Gaia XP is not assigned in the final PHAROS segmentation.

\subsubsection{\texorpdfstring{$\tau$~Bo\"otis~A}{tau Bootis A}}

$\tau$~Bo\"otis~A (HD~120136) is a bright F-type planet-host star included in PHAROS as an active, intermediate-age irradiation field. The system is particularly relevant because the primary hosts a short-period giant planet. In the SED construction, the spectrum is treated as representative of the A component, with the high-energy normalisation anchored directly by the available \textit{Chandra} grating observations.

The X-ray component is based on the \textit{Chandra}/LETG-HRC-S observations ObsIDs~17715, 20019, and 20020. The pipeline PHA2 products were split into the first-order $m=-1$ and $m=+1$ spectra for each ObsID, and background spectra were generated from the upper and lower grating background regions. The six first-order spectra were then fitted jointly in XSPEC/PyXspec over 0.2--2.0\,keV, preserving the individual ARF and RMF responses rather than combining the observations. We used C-statistics with $N_{\rm H}=10^{-4}\times10^{22}$\,cm$^{-2}$ fixed. A one-temperature model gives $C/{\rm dof}=29683.3/26775$, while the adopted two-temperature model, \texttt{tbabs*(apec+apec)}, gives $C/{\rm dof}=29668.8/26773$. The fitted temperatures are $kT_1=0.200$\,keV and $kT_2=0.463$\,keV, with a linked abundance of $Z=0.198\,Z_\odot$ and APEC normalisations of $8.12\times10^{-4}$ and $4.10\times10^{-3}$, respectively. Although the statistical improvement over the 1T model is modest, the integrated luminosity is stable between the two models; we therefore adopt the 2T model as a compact phenomenological representation of the coronal emission. The adopted unabsorbed luminosity is $L_X(0.2$--$2.0\,{\rm keV})=9.21\times10^{28}$\,erg~s$^{-1}$, with $L_X(0.1$--$2.4\,{\rm keV})=1.14\times10^{29}$\,erg~s$^{-1}$.

The adopted 2T APEC model was exported as the stellar-surface X-ray segment from 0.5166--12.3984\,nm and used as the luminosity anchor for the EUV reconstruction. The unobserved EUV region was reconstructed with the Namekata/Sanz procedure: the Namekata spectral shape was scaled so that the 12.4--91.2\,nm luminosity matches the Sanz-Forcada prediction inferred from the Chandra $L_X$, and the same scale factor was then carried to the first observed UV point. The final stitched spectrum uses the Namekata/Sanz reconstruction from 12.4--114.5\,nm, IUE/SWP gap-fill segments around 115--119\,nm and 125--198\,nm, \textit{HST}/STIS G140M from 119.5--124.9\,nm, IUE/LWP gap-fill segments through the near-UV, \textit{HST}/STIS E230H from 256.9--284.5\,nm, a short UV--Gaia bridge at 334.8--336.0\,nm, Gaia DR3 XP from 336--1020\,nm, and a PHOENIX photospheric extension from 1020.05--5000\,nm. The PHAROS product also includes an analytic Namekata/Sanz uncertainty envelope, applied at full weight over 12.4--91.2\,nm.

\subsubsection{\texorpdfstring{$\beta$~Comae~Berenices}{beta Comae Berenices}}

$\beta$~Comae~Berenices (HD~114710) is a nearby solar analogue and is treated in PHAROS as a single-star irradiation field, without a component-separation caveat. The X-ray anchor is based on the \textit{XMM-Newton}/EPIC-pn PPS spectrum from ObsID~0148680101, source~0001, using the PNS003 source spectrum together with the corresponding background spectrum, ARF, and RMF. The source is comparatively faint in the PHAROS high-energy sample, so the adopted coronal model should be interpreted mainly as an X-ray luminosity anchor for the EUV normalisation.

The EPIC-pn spectrum was fitted over 0.2--2.0\,keV in XSPEC/PyXspec using C-statistics and absorbed APEC plasma models, with $N_{\rm H}=10^{-4}\times10^{22}$\,cm$^{-2}$ fixed. A one-temperature model, \texttt{tbabs*apec}, gives $C/{\rm dof}=481.23/38$, with $kT=0.338$\,keV and an unabsorbed luminosity $L_X(0.2$--$2.0\,{\rm keV})=1.11\times10^{28}$\,erg~s$^{-1}$. We therefore adopted a two-temperature model, \texttt{tbabs*(apec+apec)}, which improves the fit to $C/{\rm dof}=370.92/36$. The adopted temperatures are $kT_1=0.162$\,keV and $kT_2=0.388$\,keV, with XSPEC normalisations of $1.49\times10^{-4}$ and $5.17\times10^{-4}$, respectively. The abundances were kept fixed at the solar value in both APEC components. The corresponding unabsorbed luminosities are $L_X(0.2$--$2.0\,{\rm keV})=1.17\times10^{28}$\,erg~s$^{-1}$, $L_X(0.2$--$2.4~{\rm keV})=1.18\times10^{28}$\,erg~s$^{-1}$, and $L_X(0.1$--$2.4\,{\rm keV})=1.34\times10^{28}$\,erg~s$^{-1}$.

The adopted XMM/APEC model was exported as the stellar-surface X-ray segment over the short-wavelength anchor, with the XMM/Namekata stitch using the APEC model to 12.4\,nm. The Sanz-Forcada normalisation uses $L_X(0.2$--$2.0~{\rm keV})=1.17\times10^{28}$\,erg~s$^{-1}$, giving $L_{\rm EUV}=8.71\times10^{28}$\,erg~s$^{-1}$. The Namekata spectrum was scaled with a wavelength-dependent normalisation: the UV-overlap scale is 6.72, the total EUV scale is 2.16, and the Sanz relative scale is 0.321, yielding a final Namekata/Sanz luminosity ratio of unity. The final stitch uses the XMM/APEC anchor from $\simeq0.527$--12.4\,nm, the Namekata/Sanz reconstruction from 12.4--115.06\,nm, HST/STIS E140M from 115.06--173.40\,nm, an IUE/SWP bridge from 173.56--198.00\,nm, IUE/LWP+LWR bridging from 198.01--219.99\,nm and 320.01--330.00\,nm, HST/STIS E230 from 220.00--320.00\,nm, Gaia DR3 XP from 330.00--1049.92\,nm, and a PHOENIX photospheric extension from 1050\,nm to 5\,$\mu$m. Thus, the final $\beta$~Com SED is suitable for PHAROS as a nearby solar-analogue irradiation field, with the main interpretation caveat that the EUV interval remains a semiempirical Namekata/Sanz reconstruction normalised by a weak but usable XMM-derived coronal anchor.

\subsubsection{70~Ophiuchi~A}

70~Ophiuchi~A (HD~165341~A) is part of a nearby visual binary, and the PHAROS spectrum constructed here is based on a component-centred Chandra/LETG extraction, likely associated with the A component. In ObsID~4482, the PHA2 extraction centre coincides with the brighter zero-order source identified by \texttt{wavdetect}, with an offset of only $\simeq0.37$ detector pixel from Source~2. We therefore treat the high-energy spectrum as a component-centred LETG product while retaining the interpretation caveat that any residual contribution from the companion depends on the detailed grating geometry.

The Chandra spectrum was represented with a two-temperature APEC coronal model fitted over 5--100\,\AA\ (0.5--10\,nm), using the $\pm1$ LETG spectra with multiorder response contributions. The adopted temperatures are $kT_1=0.148$\,keV and $kT_2=0.397$\,keV, with XSPEC normalisations of $4.08\times10^{-4}$ and $7.77\times10^{-4}$, respectively. The model gives an observed 5--100\,\AA\ flux of $3.70\times10^{-12}$\,erg~cm$^{-2}$~s$^{-1}$ and an adopted luminosity $L_X(5$--$100~\mathrm{\AA})=1.16\times10^{28}$\,erg~s$^{-1}$
for the adopted distance of $d=5.11$\,pc. 
This luminosity sets the Sanz-Forcada EUV normalisation, giving $L_{\rm EUV}(100$--$920\,\mathrm{\AA})=8.59\times10^{28}$\,erg~s$^{-1}$.

For catalogue and plotting consistency, the final SED carries the Chandra/LETG 2T APEC model to the common X-ray/EUV boundary at 12.4\,nm. The stitch then uses a residual Namekata/Sanz EUV segment from 12.4--91.2\,nm, a smooth Namekata transition to the HST UV scale from 91.2--115\,nm, HST/COS G130M from 115.0--145.5\,nm, HST/COS G160M from 145.5--171.0\,nm, HST/STIS E140M from 171.03--172.00\,nm, and HST/STIS E230M over the NUV interval up to 300\,nm, with its small internal gap retained explicitly. The optical--IR portion uses a short PHOENIX bridge from 300.1--346.5\,nm, the empirical CFLIB HD~165341 spectrum from 346.5--800.0\,nm, and a PHOENIX red extension from 800\,nm to 5\,$\mu$m. The Chandra contribution over 10--12.4\,nm is accounted for separately in the residual EUV normalisation: $L_{\rm Chandra}(10$--$12.4\,\mathrm{nm})=4.19\times10^{26}$\,erg~s$^{-1}$ and $L_{\rm Namekata}(12.4$--$91.2\,\mathrm{nm})=8.55\times10^{28}$\,erg~s$^{-1}$, giving a combined 10--91.2\,nm luminosity of $8.59\times10^{28}$\,erg~s$^{-1}$ and a final Namekata/Sanz ratio of unity.

\subsubsection{36~Ophiuchi~B}

36~Ophiuchi~B (HD~155885) is one component of the nearby 36~Oph visual binary, a pair of early-K dwarfs for which \textit{Chandra} can provide component-resolved high-energy constraints \citep{Wood.and.Linsky.2006ApJ...643..444W}. In ObsID~4483, the LETG/HRC-S observation was treated explicitly as a resolved binary dataset: 36~Oph~A and B were assigned separate zero-order sources, with 36~Oph~B corresponding to \texttt{TG\_SRCID}=2. The $\pm1$ LETG arms were extracted and modelled simultaneously. The beginning of the observation showed an elevated count-rate interval, so the first $\simeq8$~ks were excluded and the PHAROS X-ray anchor was derived from the remaining quiescent exposure of 69.512\,ks. The quiescent 36~Oph~B spectrum retains 40,254 first-order source events, split into 19,801 events in $m=-1$ and 20,453 events in $m=+1$. 

The A component was reduced and fitted in the same way and provides an important internal consistency check. Its quiescent 3T LETG model gives $L_X(0.10$--$2.40\,{\rm keV})=8.04\times10^{28}$~erg~s$^{-1}$, only $\simeq3\%$ higher than the B-component value of $7.80\times10^{28}$~erg~s$^{-1}$. This agreement supports the interpretation that the two early-K components have comparable high-energy outputs. Nevertheless, the strictly A-only panchromatic reconstruction was not adopted as the preferred PHAROS product for this binary because its UV/NUV coverage is more fragmented: the HST/STIS spectrum is split by substantial gaps, and the available IUE/SWP data do not provide a stable bridge to the optical. We therefore use 36~Oph~B as the preferred component-resolved PHAROS SED.

The 36~Oph~B LETG spectrum was fitted over 0.10--2.40\,keV in XSPEC/PyXspec using \texttt{tbabs*(apec+apec+apec)} with C-statistics, Wilms abundances \citep{Wilms.et.al.2000ApJ...542..914W}, and Verner photoionisation cross sections \citep{Verner.et.al.1996ApJ...465..487V}. The absorbing column was fixed to $N_{\rm H}=10^{-4}\times10^{22}$\,cm$^{-2}$, appropriate for a nearby source. The redshifts were fixed to zero, and a fixed global abundance of $Z=0.05\,Z_\odot$ was adopted.  
This low abundance should not be interpreted as a measured coronal abundance, but as a fitting choice that stabilises the LETG spectral shape and the integrated X-ray luminosity.
A 3T model was adopted because it improved the quiescent fit relative to the best 2T fixed-abundance solution, from $C/{\rm dof}=1.425$ to $1.314$, while changing $F_X(0.20$--$2.00\,{\rm keV})$ by only $\simeq2.1\%$. The adopted parameters are $kT_1=0.030$\,keV, $kT_2=0.239$\,keV, and $kT_3=1.666$\,keV, with APEC normalisations of $3.248\times10^{-2}$, $1.984\times10^{-2}$, and $1.043\times10^{-2}$, respectively. The fit statistic is $C=2.4963\times10^4$ for 19002 degrees of freedom. The corresponding model fluxes are $F_X(0.10$--$2.40\,{\rm keV})=1.8161\times10^{-11}$, $F_X(0.20$--$2.00\,{\rm keV})=1.2367\times10^{-11}$, and $F_X(0.20$--$2.40\,{\rm keV})=1.2926\times10^{-11}$~erg~cm$^{-2}$~s$^{-1}$, giving $L_X=7.7966\times10^{28}$, $5.3091\times10^{28}$, and $5.5492\times10^{28}$~erg~s$^{-1}$ in the same bands for $d=5.99$\,pc. The wavelength-space export over 0.5166--12.3984\,nm preserves the XSPEC luminosity, recovering $L_X(0.10$--$2.40\,{\rm keV})=7.7966\times10^{28}$~erg~s$^{-1}$ to numerical precision.

The X-ray segment was therefore carried into the PHAROS stitch as the component-resolved Chandra/LETG 3T APEC surface-flux spectrum over 0.518--12.397\,nm. Using the Sanz-Forcada relation, the adopted X-ray luminosity implies $L_{\rm EUV}(100$--$920\,{\rm \AA})=4.4365\times10^{29}$\,erg~s$^{-1}$. The Namekata-based XUV/EUV shape was scaled to this luminosity, with a raw Namekata luminosity of $4.0572\times10^{28}$~erg~s$^{-1}$ over 12.398--92\,nm, a multiplicative scale factor of 10.935, and a final Namekata/Sanz luminosity ratio of unity. In the final proxy-assisted B stitch, the Namekata/Sanz reconstruction connects the Chandra endpoint to the short-wavelength UV; the IUE/SWP and IUE/LWR B spectra provide the FUV/NUV bridge after smooth calibration using the 36~Oph~A HST/STIS spectrum as a shape proxy; Gaia DR3 XP supplies the resolved optical segment from 336--1020\,nm; and PHOENIX is used only as the red extension beyond the Gaia range.

\subsubsection{18~Scorpii}

18~Scorpii (HD~146233) is included in PHAROS as a solar-twin irradiation field, with no component-resolution caveat applied to the final SED. The high-energy anchor is based on the \textit{XMM-Newton}/EPIC-pn PPS spectrum from ObsID~0303660101, using the PNS003 source product for source~0003. Because no suitable redistribution matrix was present among the filtered PPS products, an EPIC-pn RMF was generated with \texttt{rmfgen} and associated with the adopted source spectrum, background, and ARF. The spectrum was fitted in XSPEC/PyXspec over 0.2--2.0\,keV with absorbed APEC plasma models, adopting \texttt{tbabs} absorption with $N_{\rm H}=10^{-4}\times10^{22}$\,cm$^{-2}$ fixed, solar abundance fixed, and zero redshift. A freely fitted two-temperature model formally reduced the statistic, but its hot component was unconstrained and non-physical for a quiescent solar-type corona ($kT_2\simeq64$\,keV). 
We therefore adopted the single-temperature model, \texttt{tbabs*apec}, as the more stable coronal representation for the X-ray anchor.
The adopted fit gives $kT=0.202$\,keV, APEC normalisation $1.43\times10^{-5}$, and $C/{\rm dof}=16.22/13$. The corresponding fluxes are $F_X(0.2$--$2.0\,{\rm keV})=2.27\times10^{-14}$\,erg\,cm$^{-2}$\,s$^{-1}$ and $F_X(0.1$--$2.4\,{\rm keV})=2.62\times10^{-14}$\,erg\,cm$^{-2}$\,s$^{-1}$, giving $L_X(0.2$--$2.0\,{\rm keV})=5.41\times10^{26}$\,erg\,s$^{-1}$ and $L_X(0.1$--$2.4\,{\rm keV})=6.23\times10^{26}$\,erg\,s$^{-1}$ for the adopted distance of 14.1\,pc.

The fitted APEC model was exported as an unabsorbed stellar-surface $F_\lambda$ spectrum and used as the X-ray segment. The adopted $L_X(0.2$--$2.0\,{\rm keV})$ implies $L_{\rm EUV}=6.17\times10^{27}$\,erg\,s$^{-1}$ over 100--920\,\AA\ from the Sanz-Forcada relation. The Namekata solar-scaling XUV reconstruction was then normalised so that the integrated 12.4--91.2\,nm luminosity matches this Sanz-Forcada value. The UV-overlap-scaled Namekata spectrum initially exceeded the Sanz normalisation by a factor of 5.56, and the final exact scaling sets the Namekata/Sanz luminosity ratio to unity. The compact direct-stitch construction uses XMM/APEC from 0.52--12.4\,nm, Namekata/Sanz from 12.4--115\,nm, and the observed/hybrid spectrum longward of 115\,nm. 
In the final PHAROS FITS product, these blocks are represented as XMM/EPIC-pn 1T APEC over 0.523--12.235\,nm, Namekata/Sanz over 12.5--114.5\,nm, HST/COS over 115.000--209.939\,nm, an IUE/LWP+LWR bridge over 210.206--224.885\,nm, HST/STIS E230 over 225.152--314.780\,nm, HST/STIS G430 over 315.054--559.808\,nm, HST/STIS G750 over 560.083--650\,nm, Gaia XP over 650--950\,nm, and a PHOENIX photospheric extension from 950--5000\,nm. The G750 segment is therefore used only over its cleaner blue interval, while Gaia XP provides the empirical low-resolution optical bridge to the PHOENIX photospheric model, which is scaled to the trusted Gaia interval using a smoothed-model comparison.

The resulting 18~Sco SED is, therefore, suitable as a solar-analogue PHAROS irradiation field. 
Its main interpretation caveat is not source multiplicity, but the low-count nature of the EPIC-pn X-ray anchor and the adoption of a stable 1T APEC representation.

\subsubsection{Upsilon~Andromedae~A}

Upsilon~Andromedae~A (HD~9826, HIP~7513, GJ~61) is a bright F-type planet-host star in a wide multiple system. The known low-mass companion is an M dwarf at a projected separation of $\simeq750$\,AU, or about $55$'' on the sky \citep[e.g.,][]{Lowrance.et.al.2002ApJ...572L..79L}. This separation is much larger than the HST/STIS, IUE, and Gaia XP apertures or catalogued extractions used here when centred on the primary, and we therefore treat the UV--optical/IR SED as representative of the A component. We also checked the X-ray extraction explicitly: in the quiet EPIC-pn image, the companion lies $55.36$'' from the primary, outside the adopted $40$'' source aperture, with a $\simeq15.36$'' gap to the aperture edge. A simple aperture-photometry test gives a companion-to-primary net-count ratio of only $\simeq1.2\%$ in the 0.2--2.0\,keV quiet image, indicating that contamination of the adopted X-ray spectrum by the companion is negligible for the purposes of the present SED reconstruction.

The X-ray segment is based on the \textit{XMM-Newton}/EPIC-pn observation ObsID~0722030101. The background diagnostics show a high-background rise after the first $\simeq6$\,ks, so the adopted spectrum uses the quiet-time EPIC-pn source~0001 extraction, with an effective exposure of 5280.34~s. The spectrum was fitted over 0.2--2.4\,keV in XSPEC/PyXspec using absorbed APEC models and C-statistics, with the absorbing column fixed at $N_{\rm H}=10^{-4}\times10^{22}$\,cm$^{-2}$. A one-temperature model, \texttt{tbabs*apec}, gives $C/{\rm dof}=307.62/439$, with $kT=0.260$\,keV and an APEC normalisation of $2.12\times10^{-4}$. The corresponding unabsorbed flux is $F_X(0.2$--$2.0\,{\rm keV})=3.48\times10^{-13}$\,erg~cm$^{-2}$~s$^{-1}$, giving $L_X(0.2$--$2.0\,{\rm keV})=7.56\times10^{27}$\,erg~s$^{-1}$ at $d=13.4706$\,pc; over 0.2--2.4\,keV the luminosity is essentially unchanged, $L_X=7.56\times10^{27}$\,erg~s$^{-1}$. 
A two-temperature model improves the formal statistic to $C/{\rm dof}=291.68/437$, with $kT_1=0.101$\,keV and $kT_2=0.303$\,keV, but the final PHAROS product adopts the simpler 1T model as a conservative quiet-time coronal anchor rather than interpreting the additional cool component physically.

The adopted 1T APEC model was exported as an unabsorbed stellar-surface flux segment from 0.520--12.392\,nm. Its luminosity sets the Sanz-Forcada EUV normalisation, giving $L_{\rm EUV}(100$--$920\,\mathrm{\AA})=5.96\times10^{28}$\,erg~s$^{-1}$. The Namekata-based EUV shape was then scaled to this luminosity over 12.4--91.2\,nm, reducing the initial Namekata/Sanz ratio from 11.46 to unity with a scale factor of $8.73\times10^{-2}$; the scaled 91.2--115\,nm Namekata tail is retained only for continuity with the observed UV. The stitched SED therefore uses the XMM/APEC segment from 0.520--12.392\,nm, the Sanz-normalised Namekata reconstruction from 12.4--91.2\,nm, the scaled Namekata tail from 91.2--115\,nm, HST/STIS E140M from 115.0--171.0\,nm, an IUE/SWP bridge from 171.05--185.30\,nm, and an empirically rescaled IUE/LWP+LWR bridge from 185.48--256.73\,nm. This latter bridge is mildly tilted and binned to maintain continuity between the SWP end and the STIS/E230H anchor while preserving the native IUE morphology. HST/STIS E230H anchors 256.88--284.54\,nm, PHOENIX fills the near-UV photospheric gap from 284.55--335.95\,nm, Gaia DR3 XP is used from 336--1010\,nm, and PHOENIX provides the red extension from 1020.02\,nm to 5\,$\mu$m. No NGSL segment is used for this target. The resulting SED is consequently suitable for PHAROS as a panchromatic irradiation field for Upsilon~Andromedae~A, with the interpretation caveat that the high-energy segment represents a single-epoch quiet-state coronal anchor rather than a time-variable characterisation of the stellar corona.

\subsubsection{\texorpdfstring{$\alpha$~Centauri~A}{alpha Centauri A}}

$\alpha$~Centauri~A (HD~128620, HIP~71683, GJ~559~A) is the nearby G2V primary of the spatially resolved $\alpha$~Centauri binary system. Multiple \textit{Chandra}/LETG HRC-S observations are available, sampling different coronal states. For the PHAROS quiet-state product, we adopt ObsID~7432, obtained in 2007, rather than ObsID~29 from 1999, when the A component was comparatively brighter and hotter in X-rays. This choice is consistent with the extended low-activity state of $\alpha$~Cen~A reported from long-term X-ray monitoring \citep{Ayres.2009ApJ...696.1931A,Robrade.et.al.2012A&A...543A..84R}. The LETG zero-order image resolves the A and B components, but the dispersed traces required explicit treatment of the extraction geometry. The source region was centred on the A component, and the upper background regions were displaced away from the trace of $\alpha$~Cen~B before the final PHA2 extraction. A 2100-s-binned zero-order light curve is consistent with a constant count rate, with no evidence for a flare requiring additional temporal filtering; the full 117.1\,ks exposure was therefore retained. The identification of the 2007 epoch as a low-state observation is thus supported by the published activity-cycle record rather than inferred solely from the raw count rates.

The negative- and positive-order spectra were fitted jointly over 0.2--2.0\,keV in XSPEC/PyXspec using Poisson source and background spectra. With the measured background spectra loaded, XSPEC evaluates the \texttt{cstat} fit using the W-statistic. Because the LETG/HRC-S detector does not intrinsically separate overlapping diffraction orders, response contributions from orders 1--8 were included for each dispersed arm and linked to the same physical plasma model. The absorbing column was fixed at $N_{\rm H}=10^{-4}\times10^{22}$\,cm$^{-2}$, using Wilms abundances and Verner photoelectric cross sections \citep{Wilms.et.al.2000ApJ...542..914W,Verner.et.al.1996ApJ...465..487V}. A one-temperature model, \texttt{tbabs*apec}, gives $C/{\rm dof}=10069.32/8734$, with $kT=0.0937^{+0.0020}_{-0.0019}$\,keV and an APEC normalisation of $(1.298^{+0.081}_{-0.079})\times10^{-3}$, where the uncertainties are 90\% confidence intervals. A two-temperature model gives $C/{\rm dof}=10066.32/8732$, improving the statistic by only $\Delta C=3.00$ for two additional free parameters; the corresponding AIC is slightly larger than for the 1T model. We therefore adopt the simpler 1T representation. The resulting unabsorbed luminosities are $L_X(0.2$--$2.0\,{\rm keV})=2.13\times10^{26}$\,erg~s$^{-1}$ and $L_X(0.1$--$2.4\,{\rm keV})=5.01\times10^{26}$\,erg~s$^{-1}$.

The adopted APEC model was exported as the stellar-surface X-ray segment over 0.5166--12.3984\,nm. The unobserved EUV region was reconstructed using the arithmetic mean of 24 daily FISM2 solar-minimum spectra as the spectral template. This template was scaled by a single multiplicative factor so that its integrated luminosity over 12.4--91.2\,nm matches the Sanz-Forcada prediction derived from the Chandra X-ray anchor,
$L_{\rm EUV}=5.78\times10^{27}$\,erg~s$^{-1}$. The same
FISM2/Sanz scale is retained from 91.2 to 120.0\,nm, without an
additional endpoint normalisation or wavelength-dependent tilt.
From 120.0\,nm onward, \textit{HST} spectroscopy is given priority, with \textit{IUE} retained only in intervals not covered by \textit{HST}.

Because $\alpha$~Cen~A is too bright for a directly usable Gaia DR3 XP spectrum, the 336--1000\,nm photospheric region is represented by the empirical Gaia XP spectrum of HD~30708, selected as a close spectroscopic match. Synthetic photometry of the HD~30708 XP spectrum was computed separately in the $B$, $V$, $R_{\rm C}$, and $I_{\rm C}$ bands and compared with the corresponding published photometry of $\alpha$~Cen~A. We adopted $V=0.003$, $B-V=0.633$, $V-R_{\rm C}=0.362$, and $R_{\rm C}-I_{\rm C}=0.331$ from the component-resolved bright-star photometry of \citet{Bessell.1983PASP...95..480B}. These measurements correspond to $B=0.636$, $R_{\rm C}=-0.359$, and $I_{\rm C}=-0.690$, with $V-I_{\rm C}=(V-R_{\rm C})+(R_{\rm C}-I_{\rm C})=0.693$. Synthetic magnitudes were evaluated using the Bessell Johnson--Cousins passbands \citep{Bessell.1990PASP..102.1181B}. A single multiplicative scale factor, constant with wavelength, was obtained from a joint weighted fit to the four bands and applied uniformly to the entire XP spectrum. This procedure preserves the empirical XP spectral shape but should not be interpreted as direct Gaia spectroscopy of the target. A PHOENIX atmosphere scaled to the
Gaia-twin overlap supplies the 1000--5000\,nm extension. The final product also includes an analytic FISM2/Sanz uncertainty envelope. A symmetric log-normal dispersion of 0.50\,dex is applied at full weight over 12.4--91.2\,nm and is tapered smoothly to zero between 91.2 and 115.0\,nm.

\subsubsection{\texorpdfstring{$\lambda$}{lambda}~Serpentis}

$\lambda$~Serpentis (HD~141004, HIP~77257) is a bright nearby G0V planet-host star reconstructed here as a mature solar-analogue irradiation field. The high-energy anchor is not based on a PHAROS XSPEC/APEC spectral fit. Instead, $\lambda$~Ser was observed with \textit{Chandra}/HRC-I in ObsID~22307 on 2020 April 25 for a net exposure of 6103\,s. Because HRC-I has essentially no intrinsic spectral resolution in imaging mode, the published analysis converted the measured count rate, $0.049\pm0.003$\,count~s$^{-1}$, into an unabsorbed 0.1--2.4\,keV flux using coronal emission-measure distributions and WebPIMMS response calculations. No significant variability was detected in the source events. The resulting flux is $f_X=(4.0\pm0.4)\times10^{-13}$\,erg~cm$^{-2}$~s$^{-1}$, corresponding to $L_X=(6.8\pm0.6)\times10^{27}$\,erg~s$^{-1}$, or $\log L_X=27.83$ \citep{Metcalfe.et.al.2023AJ....166..167M}. Thus, no fitted $kT$, APEC normalisation, or C-statistic is quoted for this target; the PHAROS X-ray treatment is a template-normalised high-energy segment anchored to the published luminosity.

For the final PHAROS product, the 0.55--12.35\,nm soft-X/X-ray segment was represented by the FISM2 strict solar-minimum spectral shape and scaled to the adopted published $L_X$. The unobserved EUV was then reconstructed with the same FISM2 solar-minimum template, normalised through the Sanz-Forcada relation. Using $L_X=6.8\times10^{27}$\,erg~s$^{-1}$ gives $\log L_{\rm EUV}=28.73596$ and $L_{\rm EUV}=5.44\times10^{28}$\,erg~s$^{-1}$ in the PHAROS 12.4--91.2\,nm normalisation band. The underlying FISM2 strict-minimum template, constructed from 18 daily spectra between 2019 January 15 and 2020 June 15, has $L_{\rm EUV}=5.45\times10^{27}$\,erg~s$^{-1}$ over the same band, requiring a multiplicative scale factor of 9.982. The final packaged spectrum gives $L_{0.5-12.4\,{\rm nm}}/L_X=1.000$ and $L_{12.4-91.2\,{\rm nm}}/L_{\rm EUV,Sanz}=1.000$. The FISM2/Sanz EUV core covers 12.45--91.15\,nm, while the 91.25--115.01\,nm interval is retained only as a transition bridge towards the observed UV scale and is not included in the Sanz luminosity sanity check.

The observed UV is anchored by HST/COS, with a short blue block from 115.01--119.26\,nm and a main UV block from 126.84--214.99\,nm. The internal HST gap at 119.26--126.84\,nm is filled with locally normalised IUE data, using edge scale factors of 0.0218 and 0.0601 on the blue and red sides, respectively. The redward IUE bridge covers 215.26--334.63\,nm and was scaled to the HST overlap over 205.0--214.5\,nm by a factor of 1.0459. The optical/IR portion uses a PHOENIX bridge from 334.78--349.93\,nm, scaled to the HST/IUE UV edge by a factor of 0.9287, the empirical CFLIB spectrum from 350.02--889.98\,nm scaled to PHOENIX by a factor of $8.72\times10^7$ over 500--800\,nm, and a PHOENIX red extension from 890.08\,nm to 5000\,nm. No Gaia or NGSL absolute optical anchor was used in this target-specific product, so the optical continuum should be interpreted as a PHOENIX-calibrated empirical/model hybrid rather than as a Gaia-anchored spectrum. The final $\lambda$~Ser SED is therefore suitable for PHAROS photochemical applications as a mature solar-like irradiation field, with the caveat that its high-energy region is a published-$L_X$-normalised FISM2/Sanz reconstruction rather than a directly fitted coronal spectrum.

\subsubsection{10~Tauri}

10~Tauri (10~Tau, HD~22484, HIP~16852, HR~1101) is a nearby F9IV--V star included in PHAROS as a mature, low-activity irradiation field. The available \textit{XMM-Newton} observations were inspected before selecting the high-energy anchor. We adopt ObsID~0134540601; the alternative
ObsID~0129350201 was not used because the EPIC-pn Thin1 exposure is affected by optical loading and the strictly filtered Medium-filter exposure does not provide a sufficiently useful independent spectrum. For the adopted observation, the EPIC-MOS2 net signal is consistent with zero and was excluded from the physical fit, leaving the EPIC-pn and EPIC-MOS1 spectra as the usable high-energy constraints.

The pn and MOS1 source and background spectra were fitted jointly over 0.2--2.0\,keV in XSPEC/PyXspec. The spectra were grouped to require at least five counts per background bin, and the fit used \texttt{cstat}; with Poisson background spectra attached, XSPEC evaluates the corresponding W-statistic. We used the \texttt{tbabs*apec} model with $N_{\rm H}=10^{-4}\times10^{22}$\,cm$^{-2}$ fixed, Wilms abundances, Verner photoelectric cross sections, solar coronal abundance, and zero redshift. Because of the limited net counts, only a one-temperature model was considered. The adopted fit gives $C/{\rm dof}=94.61/85$, with
$kT=0.110^{+0.024}_{-0.020}$\,keV and an APEC normalisation of $(9.8^{+8.1}_{-4.1})\times10^{-6}$, where the quoted parameter intervals correspond to 68\% confidence. The resulting unabsorbed 0.2--2.0\,keV flux is $9.59\times10^{-15}$\,erg~cm$^{-2}$~s$^{-1}$, corresponding to $L_X=2.22\times10^{26}$\,erg~s$^{-1}$. This 1T model should be interpreted as a compact phenomenological description of the weak coronal emission rather than as a detailed thermal decomposition.

The adopted XMM/APEC model was exported as the stellar-surface X-ray segment over 0.5166--12.4\,nm. The unobserved EUV region was reconstructed using the arithmetic mean of 24 daily FISM2 strict-solar-minimum spectra. The template was scaled by a single multiplicative factor so that its integrated luminosity over 12.4--91.2\,nm matches the Sanz-Forcada prediction derived from the measured XMM luminosity, $L_{\rm EUV}=2.86\times10^{27}$\,erg~s$^{-1}$. The same FISM2/Sanz scale is retained from 91.2\,nm to the first reliable IUE point at 128.3\,nm, without an additional endpoint normalisation, wavelength-dependent tilt, or synthetic transition function. The observed UV segment is provided by the IUE NEWSIPS spectrum SWP40060, restricted to quality-zero data and rebinned to 0.5\,nm. Its previously derived dual-anchor correction, constrained by the FISM2 blue edge and the HST/STIS NGSL overlap, is retained from 128.3\,nm to the adopted IUE/NGSL join at 182.3\,nm. The original HST/STIS NGSL spectrum then provides the observed UV--optical continuum to 1019.9\,nm. A PHOENIX model at $T_{\rm eff}=6000$\,K and solar metallicity, scaled multiplicatively to the NGSL red overlap, supplies the photospheric extension to 5000\,nm. The final product includes an analytic FISM2/Sanz uncertainty envelope with a 0.50-dex log-normal prediction scatter applied at full weight over 12.4--91.2\,nm and tapered smoothly to zero between 91.2 and 115.0\,nm.

\subsubsection{47~Ursae~Majoris}

47~Ursae~Majoris (47~UMa, HD~95128) is a nearby G-type planet-host star reconstructed here as a low-activity solar-analogue irradiation field. The available high-energy data were inspected using the \textit{XMM-Newton}/EPIC-pn PPS products from ObsID~0304203401. Several pn source candidates were processed, with response matrices generated using SAS/\texttt{rmfgen}. The source closest to the pointing centre is pn source~0002, whereas the brighter pn source~0001 is much farther off axis; we therefore use source~0002 as the target diagnostic rather than adopting the field-brightest source. The source~0002 spectrum contains only $\simeq222$ counts and was fitted over 0.2--2.4\,keV in XSPEC/PyXspec using absorbed APEC models, with $N_{\rm H}=10^{-4}\times10^{22}$\,cm$^{-2}$ fixed, Wilms abundances, and Verner cross sections. The free 1T model gives $C/{\rm dof}=3.85/4$, but drives the plasma temperature to $kT=3.21$\,keV, with normalisation $4.39\times10^{-5}$ and $L_X(0.2$--$2.0,{\rm keV})=8.69\times10^{26}$\,erg~s$^{-1}$. The free 2T model gives $C/{\rm dof}=0.57/2$, with $kT_1=0.118$\,keV and $kT_2=2.38$\,keV, normalisations of $1.18\times10^{-5}$ and $3.51\times10^{-5}$, and $L_X(0.2$--$2.0,{\rm keV})=1.02\times10^{27}$\,erg~s$^{-1}$. These fits are useful for source-quality assessment, but the low number of degrees of freedom and the hard fitted component make the raw XMM/APEC luminosity unsuitable as the absolute high-energy normalisation.
A constrained diagnostic model with $kT_1=0.25$\,keV and $kT_2=0.80$\,keV gives $C/{\rm dof}=41.57/4$ and $L_X(0.2$--$2.0\,{\rm keV})=6.78\times10^{26}$\,erg~s$^{-1}$; it is retained only as a source-quality diagnostic and is not used to set the final PHAROS high-energy normalisation.
These XMM-derived luminosities are more than an order of magnitude above the low-activity literature value
\citep[$\log L_X=25.45$,][]{Sanz-Forcada.et.al.2011A&A...532A...6S}, reinforcing our choice not to use the raw XMM/APEC luminosity as the final PHAROS high-energy normalisation.

The adopted PHAROS high-energy segment for 47~UMa is therefore a template-normalised reconstruction rather than a direct XMM/APEC spectral measurement. We used the FISM2 solar-minimum spectral shape \citep{Chamberlin.et.al.2020SpWea..1802588C}, normalised to the literature high-energy luminosities reported by \cite{Sanz-Forcada.et.al.2011A&A...532A...6S}. Specifically, we adopted $\log L_X=25.45$ and $\log L_{\rm EUV}=26.56$, corresponding to $L_X=2.82\times10^{25}$\,erg~s$^{-1}$ and $L_{\rm EUV}=3.63\times10^{26}$\,erg~s$^{-1}$. In the final stitched SED, the FISM2/$L_X$ segment covers 0.55--12.35\,nm, the FISM2/Sanz EUV reconstruction covers 12.45--91.15\,nm, and a FISM2 transition bridge scaled to the HST FUV level covers 91.25--114.95\,nm. The FUV/NUV data are then supplied by HST/COS from 115.00--143.92\,nm and HST/STIS from 256.91--284.55\,nm, with IUE low-dispersion bridges from 144.06--256.67\,nm and 284.69--329.80\,nm. A PHOENIX blue bridge connects 330.00--345.95\,nm to Gaia DR3 XP spectrophotometry over 346--1020\,nm, followed by a PHOENIX red extension to 5,000\,nm. The transition bridge was scaled to the HST FUV level, the IUE bridge was scaled to the HST overlap, and the PHOENIX extension was scaled to Gaia. Thus, the final 47~UMa SED is suitable for PHAROS photochemical applications, with the interpretation caveat that the high-energy region is a literature-normalised FISM2/Sanz reconstruction supported by a weak XMM diagnostic.

\FloatBarrier
\section{Discussion}
\label{sec:discussion}

The PHAROS sample shows that stellar age provides a useful organising coordinate for panchromatic irradiation fields, but it is not by itself a deterministic predictor of the short-wavelength spectrum. The broad tendency for younger stars to exhibit stronger X-ray, EUV, and ultraviolet emission is accompanied by substantial dispersion within the age-context groups. Rotation, magnetic activity, multiplicity, coronal state, and the observing epoch can therefore produce irradiation environments that differ appreciably even for stars assigned to similar evolutionary contexts. The Hadean, Archean, and Proterozoic groupings should consequently be interpreted as comparative age bins rather than as a one-to-one empirical time sequence for the Sun.

The component-resolved construction is particularly important for atmospheric applications because the degree of empirical constraint is strongly wavelength dependent. The optical and infrared regions are generally anchored by observed photospheric continua and model extensions, whereas the EUV region remains intrinsically reconstruction dominated. In addition, the high-energy constraint is not homogeneous across the full sample: most targets use target-specific coronal plasma models, while the weakest cases rely on literature-normalised empirical templates. Downstream photochemical, atmospheric-escape, and retrieval calculations should therefore use the stitched SED together with the provenance information and, where relevant, the supplied EUV uncertainty envelope rather than treating every wavelength interval as equally well measured.

A further limitation is that the archival components entering a given SED are generally non-simultaneous. Stellar activity cycles, rotational modulation, and transient events can alter the high-energy and ultraviolet spectrum between observing epochs, so the PHAROS products should be regarded as reference irradiation fields assembled from the best available constraints rather than as contemporaneous snapshots of the stellar atmosphere. The analytic EUV uncertainty envelope captures the dominant normalisation uncertainty adopted for the reconstructed EUV segment, but it is not a complete variability or systematic-error budget. Future releases incorporating additional time-domain information and contemporaneous multiwavelength observations would provide a natural extension of the present catalogue.

\section{Conclusions}
\label{sec:conclusions}

We have presented PHAROS, the Panchromatic Habitable-world Archive of Radiation from Observed Solar-like stars, a catalogue of component-resolved panchromatic SEDs for fifteen nearby F-, G-, and K-type stars. 
The first PHAROS release provides wavelength-dependent stellar irradiation fields for studies of exoplanet atmospheres, photochemistry, climate, atmospheric escape, and habitability around solar-like stars.
The adopted stellar ages span 0.26--6.5\,Gyr and are organised for comparison using Hadean, Archean, and Proterozoic solar-age contexts, together with a group of mature stars older than the present-day Sun. These groupings facilitate comparisons across broad evolutionary intervals without treating the individual targets as exact snapshots of the Sun at those epochs.

Each PHAROS SED covers 0.5--5000\,nm and combines observational, semiempirical, empirical-template, and model components. The high-energy region is anchored by \textit{XMM-Newton} or \textit{Chandra} data when the available observations support a target-specific coronal spectral model. In those cases, the X-ray luminosity provides the normalisation anchor for the semiempirical EUV reconstruction described in Sect.~\ref{subsec:euv_reconstruction}. For targets without a robust target-specific X-ray spectral constraint, the high-energy region is instead represented by an empirical template reconstruction normalised to literature luminosity measurements, as described in Sect.~\ref{subsec:xray_modeling}. This distinction is preserved in the component and provenance information of each final SED.

A central feature of PHAROS is the preservation of component provenance throughout each stitched SED. The standardised FITS products, segment tables, and diagnostic records identify whether each wavelength interval is directly observed, reconstructed, represented by a scaled empirical template, or model-dependent. Users can therefore trace the origin of the adopted irradiation field, identify the regions dominated by reconstruction uncertainties, and evaluate how individual methodological choices affect downstream calculations.

By extending this component-resolved approach to a curated sample of solar-like stars, PHAROS provides empirically anchored irradiation fields spanning a broad range of activity states and evolutionary stages. These spectra are not intended to represent exact snapshots of the Sun at different ages, but they enable controlled comparisons of how stellar spectral shape influences atmospheric chemistry, escape, and potential biosignature interpretation. They can also serve as stellar inputs for forward models, retrieval experiments, and atmospheric-context studies in preparation for future facilities such as NASA's Habitable Worlds Observatory.

Future PHAROS releases can expand the stellar sample, incorporate additional time-domain constraints, improve the treatment of stellar variability, and update the high-energy reconstructions as new X-ray, ultraviolet, and stellar-activity measurements become available. Such extensions will further improve the use of observed solar-like stellar irradiation fields in comparative exoplanet atmosphere and habitability studies.

\begin{acknowledgements}
V. Sumida and A. Valio acknowledge the partial financial support received from Brazilian FAPESP grants \#2021/14897-9, \#2021/02120-0, and \#2024/03652-3, as well as CAPES and MackPesquisa funding agencies.

Part of this research was carried out at the Jet Propulsion Laboratory, California Institute of Technology, under a contract with the National Aeronautics and Space Administration (80NM0018D0004).

\textit{Facilities:} XMM-Newton (EPIC-pn, EPIC-MOS), Chandra (LETG/HRC-S, HRC-I), HST (COS, STIS), IUE (SWP, LWP, LWR), Gaia (XP), and KPNO:CFT.

\textit{Software:} CIAO \citep{Fruscione.et.al.2026ApJ..1005..116F}, XMM-Newton SAS \citep{Gabriel.et.al.2004ASPC..314..759G}, XSPEC/PyXspec \citep{Arnaud.1996ASPC..101...17A}, Astropy \citep{Astropy.Collaboration.2013AA...558A..33A,Astropy.Collaboration.2018AJ....156..123A,Astropy.Collaboration.2022ApJ...935..167A}, astroquery \citep{Ginsburg.et.al.2019AJ....157...98G}, NumPy \citep{Harris.et.al.2020Natur.585..357H}, pandas \citep{McKinney.2010scpy.soft.....M}, Matplotlib \citep{Hunter.2007CSE.....9...90H}, and gyro-interp \citep{Bouma.et.al.2023ApJ...947L...3B}.
\end{acknowledgements}

\bibliographystyle{bibtex/aa.bst}
\bibliography{references}

\end{document}